\pdfoutput=1

\documentclass[english,preprint2]{aastex}
\usepackage[T1]{fontenc}
\usepackage[latin9]{inputenc}
\usepackage{listings}
\usepackage{amsthm}
\usepackage{amssymb}
\usepackage{graphicx}

\makeatletter

\providecommand{\tabularnewline}{\\}

\usepackage{amsmath}

\makeatother

\usepackage{babel}
\begin{document}

\title{First- and Second-order Fermi Acceleration at Parallel Shocks}

\shorttitle{Fermi Acceleration at Parallel Shocks}
\shortauthors{Sandroos}
\author{Sandroos, A.\altaffilmark{1}}
\affil{Finnish Meteorological Institute, PO Box 503, FI-00101, Helsinki, Finland.}
\authoremail{arto.sandroos@fmi.fi}
\author{Li, G.}
\affil{Department of Physics, University of Alabama in Huntsville, Huntsville, AL 35805, USA.}
\and
\author{Zhao, L.}
\affil{Department of Physics, University of Alabama in Huntsville, Huntsville, AL 35805, USA.}
\altaffiltext{1}{Space Sciences Laboratory, University of California, Berkeley, CA 94720, USA.}
\begin{abstract}
We report on a new Monte Carlo method for simulating diffusive shock
acceleration (DSA) of solar energetic particles at upstream and downstream
regions of quasi-parallel collisionless shock waves under the influence
of self-generated turbulence. By way of example, we apply the model
to a fast 1500 km $\mathrm{s}^{-1}$ coronal mass ejection at ten
solar radii. Results indicate that the maximum energies at outer corona
are likely to be limited to few MeV, due to lack of suprathermal protons
for appreciable wave growth, and insufficient time required acceleration.
We find that the second-order Fermi acceleration, although being a
too slow process to have a notable effect at the highest energies,
significantly flattens energy spectra at low energy end. Simulations
indicate that protons continue to damp waves efficiently several solar
radii from the shock in the downstream region, which may be an important
mechanism for heating suprathermals. Our simulations also suggest
that models assuming a simple isotropic scattering are likely to predict
too efficient acceleration. 
\end{abstract}

\keywords{acceleration of particles, methods: numerical, shock waves, Sun: corona, Sun: coronal mass ejections (CMEs), Sun: particle emission}

\section{Introduction}

Acceleration of ions by collisionless shocks driven by coronal mass
ejections (CMEs) is the best developed theory for the genesis of gradual
solar energetic particle (SEP) events. The underlying mechanism is
the diffusive shock acceleration (DSA), in which ions scatter off
waves present in the upstream and downstream regions of the shock,
cross the shock multiple times, diffuse in momentum\citep{1977ICRC...11..132A,1977DoSSR.234.1306K,1978MNRAS.182..147B,1978ApJ...221L..29B}. 

Effective DSA requires that the near upstream region is highly turbulent
to provide enough scattering to keep ions close to the shock, otherwise
ions can easily escape the acceleration region to interplanetary space
without gaining much energy. In some SEP events there are enough particles
to amplify the waves near the shock, sometimes even by several orders
of magnitude \citep{2003ApJ...591..461N,2007ApJ...658..622V}, in
which case the SEP intensities end up being streaming limited \citep{1998ApJ...504.1002R}.
Shock drift acceleration (SDA) at oblique shocks can further speed
up the acceleration \citep{1983ApJ...270..537W,1987ApJ...313..842J}.

Over the years, many quantitative models have attempted to couple
wave generation with DSA by employing various simplifying assumptions.
Monte-Carlo models typically consider ion acceleration in a predescribed
turbulence \citep{1988SSRv...48..195D,1990ApJ...360..702E,2000ApJ...528.1015V,2005ApJ...624..765G,2006A&A...455..685S}.
Quasi-stationary models assume that the ions and wave spectra near
the shock will quickly reach steady-state conditions \citep{2000JGR...10525079Z,2003JGRA..108.1082L,2005JGRA..110.6104L}.
Self-consistent models exist \citep{1999GeoRL..26.2145N,2003ApJ...591..461N,2007ApJ...658..622V,2011A&A...535A..34B},
but they are limited to one spatial dimension which limits their applicability
for the solar corona, where the shock and magnetic geometry are very
complex and vary with time. Very few models include the downstream
region self-consistently \citep{2008ApJ...686L.123N}. Instead, an
analytic model is typically used, e.g., probability of return method
by \citet{1978MNRAS.182..147B}, to return a fraction of transmitted
particles back to the upstream region. However, detailed modeling
of the wave and particle intensities both upstream and downstream
of shocks is very important for the Solar Probe Plus mission, which
will hopefully provide in-situ observations of SEPs while they are
still being accelerated near coronal shock waves.

Corsair/SEP is a project aiming to develop a self-consistent model,
in the sense of conserving the total energy of ions and waves during
scattering, for DSA of SEPs at coronal shocks in multiple spatial
dimensions. This allows us to take into account effects due to variations
in local shock obliquity angle arising from, e.g., inhomogeneous coronal
magnetic field and/or curved shock surface \citep{2006ApJ...646.1319T,2006A&A...455..685S,2007ApJ...662L.127S},
to model longitudinal distribution of SEP intensities, and study how
they change over time.

The purpose of this manuscript is to introduce and validate the numerical
DSA model implemented in Corsair/SEP code. We apply it to DSA at parallel
shocks using parameters that are suitable for solar corona near ten
solar radii, and discuss the self-generated wave spectra in detail.

The manuscript is organized as follows: Section \ref{sec:Numerical-Method}
presents the numerical model in detail, divided into subsections giving
an overview of the simulation (Sections \ref{sub:Overview-Of-Simulation}-\ref{sub:Simulation-Mesh}),
model for ambient plasma and suprathermals (Section \ref{sub:Ambient-Plasma-Model}),
Alfvén waves (Section \ref{sub:Alfv=0000E9n-waves}) and their growth
(Section \ref{sub:Wave-Growth-Factors}), and ion propagation and
pitch-angle scattering (Section \ref{sub:Particles}). In Section
\ref{sec:Test-Cases} we compare the code against a few test cases.
Results for DSA at parallel shocks are presented and discussed in
Section \ref{sec:Results}. Section \ref{sec:Conclusions} ends the
manuscripts with a summary.

\section{Numerical Model\label{sec:Numerical-Method}}

\subsection{Overview Of Simulation\label{sub:Overview-Of-Simulation}}

A summary of simulation propagation loop in pseudocode is given below
for reference. The individual steps and underlying physics are documented
in Sections \ref{sub:Simulation-Mesh}-\ref{sub:Particle Splitters}.

\begin{lstlisting}[basicstyle={\small}]
for each timestep
1. Re-evaluate global time step
2. Scattering
  2.1. accumulate wave packet energies
  2.2. convert wave energy
       density to intensity
  2.3. scatter particles and 
       accumulate growth factors
  2.4. apply wave growth
3. Propagation
  3.1. propagate particles
  3.2. propagate wave packets
  3.2. apply boundary conditions
4. inject new particles
   and wave packets
5. Particle Splitting
\end{lstlisting}

\subsection{Simulation Mesh\label{sub:Simulation-Mesh}}

In Corsair/SEP waves and protons are modeled using macroparticles.
A two-dimensional $(x,\lambda)$ phase-space mesh is used to mediate
wave-particle interactions. The phase-space mesh is further split
into a one-dimensional mesh in configuration space ($x$ coordinate,
$\mathrm{i}$ index) with uniform cell size $\Delta x$ and volume
$\Delta x^{3}$. Each spatial cell contains a wavelength mesh ($\lambda$
coordinate, $\mathrm{l}$ index) with uniform cell size in logarithmic
units, $\ln\Delta\lambda=\ln\left(\lambda_{\mathrm{l+1}}-\lambda_{\mathrm{l}}\right)$,
where $\lambda_{\mathrm{l}}$ are the node coordinates. A cell having
indices $(\mathrm{i,l})$ covers the phase-space $[x_{\mathrm{i}},x_{\mathrm{i}+1})\times[\lambda_{\mathrm{l}},\lambda_{\mathrm{l}+1})$
in physical coordinates, $[\mathrm{i},\mathrm{i}+1)\times[\mathrm{l},\mathrm{l}+1)$
in logical coordinates.

We use $\lambda$ instead of the wave number $k=2\pi/\lambda$ to
simplify scattering across pitch (pitch angle cosine) $\mu=0^{\pm}$
-- ions cross $\lambda=0$ point instead of $k=\pm\infty$. A convention
that positive (negative) $\lambda$ correspond to left-handed L (right-handed
R) wave helicities is used. 

Simulation uses logical coordinates instead of physical ones. For
an ion in position $x,$ the corresponding logical coordinate is $x_{\mathrm{L}}=\left(x-x_{0}\right)/\Delta x$,
i.e., the $\mathrm{i}$-index of the spatial cell in which the particle
currently resides is given by the integer part of $x_{\mathrm{L}}.$
Logical wavelength coordinate is

\begin{equation}
\lambda_{\mathrm{L}}=\begin{cases}
N_{\lambda}/2-\max\left(0,\ln\left|\lambda/\lambda_{0}\right|\right), & \lambda<0,\\
N_{\lambda}/2+\max\left(0,\ln\lambda/\lambda_{0}\right), & \lambda\geqslant0,
\end{cases}
\end{equation}
where $N_{\lambda}$ is the (even) number of cells in wavelength mesh.
Linear units are used for the two cells neighboring $\lambda=0$ point,
$\left|\lambda\right|\leqslant\lambda_{0}$, to improve mesh resolution
at longer wavelengths.

Interpolations and accumulations between particle positions and the
phase-space mesh are carried out in logical coordinates, using particle
shape factors commonly used in particle-in-cell methods \citep[e.g.,][]{1985PlasmaPhysBirdsall,1989ComSimParHockney}.
For example, in one dimension, triangular-shaped clouds (TSC) interpolate
a scalar quantity $\phi$ from mesh to position $x_{\mathrm{L}}$
using the values $\phi_{\mathrm{i}-1},\,\phi_{\mathrm{i}},\,\phi_{\mathrm{i}+1}$
from nearest cells as $\phi(x_{\mathrm{L}})=\sum S_{\mathrm{n}}\phi_{\mathrm{n}}$,
where 

\[
S_{\mathrm{n}}=\begin{cases}
(\mathrm{i}+1-x_{\mathrm{L}})^{2}/2 & \mathrm{for\; n=}\mathrm{i}-1,\\
(x_{\mathrm{L}}-\mathrm{i})^{2}/2 & \mathrm{for\; n=}\mathrm{i}+1,\\
1-S_{\mathrm{i}-1}-S_{\mathrm{i}+1} & \mathrm{for\; n=}\mathrm{i},\\
0 & \mathrm{otherwise.}
\end{cases}
\]
In two dimensions the interpolated value is simply the product of
one-dimensional shape factors, $\phi(x_{\mathrm{L}},\lambda_{\mathrm{L}})=\sum S_{n}S_{m}\phi_{nm}.$
Henceforth we will drop the summation symbols and use a short-hand
notation $S_{\mathrm{i}}S_{\mathrm{l}}\phi_{\mathrm{il}}=S_{\mathrm{il}}\phi_{\mathrm{il}}$.

\subsection{Ambient Plasma Model\label{sub:Ambient-Plasma-Model}}

Simulations are carried out in the shock normal incidence frame. In
the upstream region (subscript 1) plasma has a constant mass density
$\mathrm{\rho_{\mathrm{m1}}}$, speed $V_{\mathrm{n1}}$, and magnetic
field $B_{\mathrm{n1}}$. Plasma parameters in the downstream region
(subscript 2) are solved using ideal MHD Rankine-Hugoniot equations,

\begin{eqnarray*}
\frac{\rho_{\mathrm{m2}}}{\rho_{m1}} & = & r,\\
\frac{V_{\mathrm{n2}}}{V_{\mathrm{n1}}} & = & r^{-1},\\
\frac{B_{\mathrm{n2}}}{B_{\mathrm{n1}}} & = & 1,\\
\frac{p_{2}}{p_{1}} & = & \frac{\left(\gamma+1\right)r-\left(\gamma-1\right)}{\left(\gamma+1\right)-\left(\gamma-1\right)r},
\end{eqnarray*}
where $\gamma=5/3$ is the polytropic index, $r$ is the gas compression
ratio, $p$ is the pressure, and subscript $\mathrm{n}$ refers to
normal component with respect to shock normal.

Suprathermal ions are assumed to have a Kappa distribution in energy
$U$,

\begin{equation}
f_{\kappa}(U)=\frac{C_{\Gamma}}{\kappa U_{\kappa}}\sqrt{\frac{U}{\kappa U_{\kappa}}}\left(1+\frac{U}{\kappa U_{\kappa}}\right)^{-(\kappa+1)},\label{eq:kappa energy distrib}
\end{equation}
where 

\begin{eqnarray}
C_{\Gamma} & = & \frac{\Gamma(\kappa+1)}{\Gamma(\kappa-\frac{1}{2})\Gamma(\frac{3}{2})},\\
U_{\mathrm{\kappa}} & = & \left(\frac{2\kappa-3}{2\kappa}\right)k_{\mathrm{B}}T.
\end{eqnarray}
Here $T$ is plasma Maxwell-Boltzmann temperature, and $\kappa U_{\mathrm{\kappa}}$
is thermal energy in kappa distribution. We assume ion distributions
to be isotropic in momentum space.

\subsection{Alfvén waves\label{sub:Alfv=0000E9n-waves}}

We model Alfvén waves propagating parallel (superscript $+)$ and
antiparallel (superscript $-$) to ambient magnetic field as monochromatic
wave packets, each carrying the total (electric + magnetic) energy
in waves over a wavelength range,

\begin{equation}
U_{\mathrm{packet}}=\frac{1}{\mu_{0}}\int d^{3}\mathbf{r}\int_{\lambda_{\mathrm{min}}}^{\lambda_{\mathrm{max}}}\delta B^{2}(\mathbf{r},\lambda)d\lambda,\label{eq:wave packet energy}
\end{equation}
where $\delta\mathbf{B}$ is the perturbed magnetic field.

Wave packets are propagated using WKB ray-tracing equations \citep[e.g.,][]{1965JFM....22..273W},

\begin{eqnarray}
\frac{d\mathbf{R_{\mathrm{L,R}}^{\pm}}}{dt} & = & \mathbf{V}_{\mathrm{w}}=\mathbf{V}_{\mathrm{p}}\pm\mathbf{V}_{\mathrm{A}},\\
\frac{d\mathrm{log}(\lambda_{\mathrm{L,R}}^{\pm})}{dt} & = & \frac{d}{dr}\left(\mathbf{V}_{\mathrm{p}}\pm\mathbf{V}_{\mathrm{A}}\right),\\
\mathbf{V}_{\mathrm{A}} & = & \mathbf{B}_{0}/\sqrt{\mu_{0}\rho_{m}},
\end{eqnarray}
where $\mathbf{V}_{\mathrm{p}}$ and $\mathbf{V}_{A}$ are the plasma
and Alfvén velocities at position $\mathbf{R}$, and $\mathbf{B}_{\mathrm{o}}$
is the ambient magnetic field. Henceforth the subscripts L and R,
and superscripts $\pm$, are dropped whenever the intention is clear.

Alfvén waves are assumed to have a power-law spectral intensity $I(k)=I_{0}(k/k_{0})^{\sigma}$
in the ambient coronal plasma. The corresponding wavelength spectrum
can be calculated using the relation

\begin{equation}
I(\lambda)d\lambda=I(k)dk=\frac{k^{2}}{2\pi}I(k)d\lambda.
\end{equation}
Thus, if the $k$-spectrum has a spectral index $\sigma$, the spectral
index of $\lambda$-spectrum is $s=\sigma+2$. Intensity is related
to total wave energy density as

\begin{equation}
U_{\mathrm{w,tot}}=\frac{\delta B^{2}}{\mu_{0}}=\frac{1}{\mu_{0}}\int I(\lambda)d\lambda.
\end{equation}

Finally, we normalize wave energy densities to energy density of ambient
magnetic field,

\begin{equation}
\frac{1}{\mu_{0}}\int I_{\mathrm{L,R}}^{\pm}(\lambda)d\lambda=C_{\mathrm{L,R}}^{\pm}\cdot\frac{B_{0}^{2}}{2\mu_{0}},
\end{equation}
where $C_{\mathrm{L,R}}^{\pm}<1$ is a constant.

Each time step, wave packet energies are accumulated to the phase-space
mesh to form wave spectral energy $U_{\mathrm{w}}(\mathbf{r},\lambda)$.
Each phase-space cell neighboring a wave packet's position receives
a value $U_{\mathrm{w,}il}=S_{il}U_{\mathrm{packet}}$. The obtained
$U_{\mathrm{w}}\left(\mathbf{r},\lambda\right)$ is then converted
into intensity in preparation for the scattering step,

\begin{equation}
I_{\mathrm{il}}=\mu_{0}U_{\mathrm{w,il}}/\left(\Delta x^{3}\Delta\lambda_{\mathrm{l}}\right).
\end{equation}

\subsubsection{Injection\label{sub:Injection}}

New wave packets are injected to the spatial cell on $-x$ inflow
boundary whenever there are no wave packets in it. Injection coordinates
are 

\begin{eqnarray}
x_{\mathrm{inj,L}} & = & \frac{1}{2}+\frac{V_{\mathrm{w}}}{\Delta x}t,\label{eq:injection x-offset}\\
\lambda_{\mathrm{inj}} & = & \frac{1}{2}\left(\lambda_{\mathrm{l}}+\lambda_{\mathrm{l}+1}\right),
\end{eqnarray}
where $t$ is the current simulation time. Injection position is clamped
to be inside the inflow cell, $x_{\mathrm{inj,L}}\in\left[0,1\right]$.
The wave packets are injected to physical cell centroids $\lambda_{\mathrm{inj}}=\lambda_{\mathrm{l}}+0.5\,\Delta\lambda$,
instead of logical $\lambda_{\mathrm{L}}=\mathrm{l}+0.5$ centroids,
as this gives interpolated intensities that better correspond to the
analytic intensities. The offset distance in Eqn. (\ref{eq:injection x-offset})
maintains a constant distance $\Delta x$ between subsequent wave
packets. Injection energy is

\begin{equation}
U_{\mathrm{packet}}=\frac{\Delta x^{3}}{\mu_{0}}\int_{\lambda_{\mathrm{l}}}^{\mathrm{\lambda_{\mathrm{l+1}}}}I(\mathrm{\lambda)d\lambda}.
\end{equation}
Wave packets hitting the $+x$ outflow boundary are removed.

\subsubsection{Shock Boundary Conditions\label{sub:Shock-Boundary-Conditions}}

Waves crossing the shock front experience a jump in wavelength because
$\mathbf{V}_{\mathrm{w}}$ in upstream and downstream regions differ.
Incident waves also split into transmitted and reflected waves, conserving
frequency and helicity. Boundary conditions discussed below are applied
to the incident wave packets whenever a shock crossing is detected
to create a transmitted wave, and a new reflected wave packet is injected
to the simulation.

Transmitted (T) and reflected (R) wavelengths are given in terms of
incident (I) wavelength as

\begin{eqnarray}
\frac{\lambda^{\mathrm{R}}}{\lambda^{\mathrm{I}}} & = & \frac{V_{\mathrm{p,2}}\mp V_{\mathrm{A,2}}-V_{\mathrm{shock}}}{V_{\mathrm{p,1}}\pm V_{\mathrm{A,1}}-V_{\mathrm{shock}}},\label{eq:reflected wavelength}\\
\frac{\lambda^{\mathrm{T}}}{\lambda^{\mathrm{I}}} & = & \frac{V_{\mathrm{p,2}}\pm V_{\mathrm{A,2}}-V_{\mathrm{shock}}}{V_{\mathrm{p,1}}\pm V_{\mathrm{A,1}}-V_{\mathrm{shock}}},\label{eq:transmitted wavelength}
\end{eqnarray}
where the upper (lower) signs hold for incident parallel (antiparallel)
propagating waves.

Reflection (R) and transmission (T) coefficients for the wave amplitude
$\delta\mathbf{B}$ are given in \citet{1992A&A...263..413C,1998A&A...331..793V}.
For wave packet energies they are

\begin{eqnarray}
R & = & \left\{ \frac{\sqrt{r}(\sqrt{r}-1)}{2}\frac{M_{1}+1}{M_{1}-\sqrt{r}}\right\} ^{2}\frac{\lambda^{\mathrm{R}}}{\lambda^{\mathrm{I}}},\label{eq:wave energy reflection coeff}\\
T & = & \left\{ \frac{\sqrt{r}(\sqrt{r}+1)}{2}\frac{M_{1}+1}{M_{1}+\sqrt{r}}\right\} ^{2}\frac{\lambda^{\mathrm{T}}}{\lambda^{\mathrm{I}}},\label{eq:wave energy transmission coeff}
\end{eqnarray}
where $M_{1}=V_{\mathrm{n1}}/V_{\mathrm{A1}}$ is the Alfvénic Mach
number in the upstream region. In Eqns. (\ref{eq:wave energy reflection coeff})
and (\ref{eq:wave energy transmission coeff}) the values in brackets
are the reflection and transmission coefficients of wave intensities
calculated in \citet{1992A&A...263..413C}. The $\lambda^{\mathrm{T,R}}/\lambda^{\mathrm{I}}$
terms stem from the fact that in downstream region the wave packets
are closer together than in upstream region. This effect is included
in the intensity coefficients, but here we need to factor it out to
avoid double-counting it.

\subsection{Particles\label{sub:Particles}}

Guiding center (GC) approximation is used for ions with state variables
($\mathbf{R},V_{\parallel},\mathcal{M}$), where $\mathcal{M}=\frac{1}{2}mv^{2}(1-\mu^{2})/B_{0}$
is the magnetic moment. GC equations of motion are \citep{1983ApJ...270..537W}

\begin{eqnarray}
\frac{d\mathrm{\mathbf{R}}}{dt} & = & \mathbf{V}_{\parallel}+\mathbf{V}_{\mathrm{D}},\\
\frac{dV_{\parallel}}{dt} & = & -\frac{\mathcal{M}}{m}\nabla B_{0,\parallel}\\
 & + & \frac{\mathbf{V}_{\mathrm{E}}}{B_{0}}\cdot(\partial_{t}+\left[\mathbf{V}_{\parallel}+\mathbf{V}_{\mathrm{E}}\right]\cdot\nabla)\mathrm{B}_{0},\\
\mathcal{M} & = & \mathrm{constant,}
\end{eqnarray}
where the drift velocity

\begin{eqnarray}
\mathbf{V}_{\mathrm{D}} & = & \mathbf{V}_{\mathrm{E}}+\mathbf{V}_{\nabla B}+\mathbf{V}_{\mathrm{C}}\\
 & = & \frac{\mathbf{E}\times\mathbf{B_{0}}}{B_{0}^{2}}+\frac{\mathbf{B}_{0}\times\nabla B_{0}}{B_{0}^{2}}\\
 & + & \frac{mV_{\parallel}^{2}}{q}\frac{\mathbf{B}\times\left(\mathbf{B}\cdot\nabla\right)\mathbf{B}}{B^{4}}
\end{eqnarray}
is the sum of electric, gradient, and curvature drifts. Electric field
is the convectional $\mathbf{E}=-\mathbf{V}_{\mathrm{p}}\times\mathbf{B}_{0}$.
Scattering also modifies ions' $\mathcal{M}$ and is handled in a
separate step (see below).

\subsubsection{Pitch-Angle Scattering\label{sub:Scattering}}

The quasilinear theory states that the pitch angle diffusion is governed
by the $\mu-$dependent part of particle transport equation \citep{1966ApJ...146..480J},

\begin{eqnarray}
\frac{\partial f}{\partial t} & = & \frac{\partial}{\partial\mu}\left(D_{\mu\mu}\frac{\partial f}{\partial\mu}\right)\label{eq:fokker-planck}\\
 & = & \frac{\partial^{2}\left(D_{\mu\mu}f\right)}{\partial\mu^{2}}-\frac{\partial}{\partial\mu}\left(\frac{\partial D_{\mu\mu}}{\partial\mu}f\right).
\end{eqnarray}
Pitch angle diffusion coefficient in parallel or antiparallel wave
rest frame $\mathbf{V}_{\mathrm{w}}=\mathbf{V}_{\mathrm{p}}\pm\mathbf{V}_{\mathrm{A}}$
(variables with tilde) is 

\begin{eqnarray}
D_{\mu\mu} & = & \frac{\pi}{2}\frac{\Omega}{B_{0}^{2}}\left(1-\tilde{\mu}^{2}\right)\left|k_{\mathrm{res}}\right|I(\mathbf{r},k_{\mathrm{res}})\label{eq:D_mumu k}\\
 & = & \frac{\pi}{2}\frac{\Omega}{B_{0}^{2}}\left(1-\tilde{\mu}^{2}\right)\left|\lambda_{\mathrm{res}}\right|I(\mathbf{r},\lambda_{\mathrm{res}}),\label{eq:D_mumu lambda}
\end{eqnarray}
evaluated at the resonant wavelength

\begin{equation}
\lambda_{\mathrm{r}}=2\pi/k_{\mathrm{r}}=\left(2\pi/\Omega\right)\tilde{V}\tilde{\mu}.\label{eq:resonant wavelength}
\end{equation}

The Fokker-Planck Eqn. (\ref{eq:fokker-planck}) is formally equivalent
to a stochastic differential equation

\begin{equation}
dX_{t}=a\, dt+b\, dW_{t},\label{eq:stochastic de}
\end{equation}
where drift $a=\partial_{\mu}D_{\mu\mu}$, variance $b^{2}=2D_{\mu\mu}$,
$W_{t}$ is a Wiener process, and $X_{\mathrm{t}}$ is the sought-after
solution (particle pitch here). We proceed by solving Eqn. (\ref{eq:stochastic de})
using a first-order accurate implicit predictor-corrector method \citep{1997NumSolSDE},

\begin{eqnarray}
\tilde{\mu}_{\mathrm{trial}} & = & \tilde{\mu}_{t}+a(\tilde{\mu}_{t})\, dt+b(\tilde{\mu}_{t})\,\sqrt{dt}\, N,\label{eq:scatter step 1}\\
a_{0} & = & \left(1-\eta\right)a(\tilde{\mu}_{t}),\label{eq:scatter step 2}\\
a_{1} & = & \left(1-\eta\right)a(\tilde{\mu}_{\mathrm{trial}}),\label{eq:scatter step 3}\\
\tilde{\mu}_{t+1} & = & \tilde{\mu}_{0}+(\theta a_{1}+\left(1-\theta\right)a_{0})dt\nonumber \\
 & + & \left(\eta b(\tilde{\mu}_{t})+\left(1-\eta\right)b(\tilde{\mu}_{\mathrm{trial}})\right)\nonumber \\
 &  & \cdot\sqrt{dt}\, N,\label{eq:scatter step 4}
\end{eqnarray}
where $\tilde{\mu}_{\mathrm{t}}$ and $\tilde{\mu}_{\mathrm{t+1}}$
are the initial and new pitch, and $N$ is a normally distributed
random number with zero mean and unit variance. We use values $\theta=0$,
$\eta=0.5$, for the impliciteness parameters. Particles scattering
outside the $\left|\tilde{\mu}\right|\leqslant1$ interval are reflected
at the boundaries, $\tilde{\mu}\rightarrow\tilde{\mu}'=1-\tilde{\mu}$.

The diffusion term $b$ is interpolated to ion phase-space position
as follows: first the intensity is interpolated in $x$ to cell centroids
neighboring the ion in $\lambda$ direction, $I_{l}=S_{i}I_{il}$.
Then diffusion terms $b_{\mathrm{L,U}}^{2}=D_{0}(1-\mu_{\mathrm{L,U}}^{2})\left|\lambda^{\mathrm{L,U}}\right|I^{\mathrm{L,U}}$
are calculated, where $D_{0}=\pi\Omega/\left(2B_{0}^{2}\right)$,
on the upper (U) and lower (L) cell faces using average intensities
$I^{\mathrm{L}}=\left(I_{l-1}+I_{l}\right)/2$, $I^{\mathrm{U}}=\left(I_{l}+I_{\mathrm{l+1}}\right)/2$.
Finally, $a$ and $b$ terms are interpolated to ion position as

\begin{eqnarray}
a & = & \frac{b_{\mathrm{\mathrm{U}}}^{\mathrm{2}}-b_{\mathrm{L}}^{2}}{2\left(\lambda_{l+1}-\lambda_{l}\right)},\label{eq:numeric drift}\\
b^{2} & = & b_{\mathrm{L}}^{2}+2a\left(\lambda_{\mathrm{r}}-\lambda_{l}\right).\label{eq:numeric diffusion}
\end{eqnarray}
Thus, $a$ is the derivative of $b$ with respect to $\lambda$.

In principle the scattering algorithm, given in Eqns. (\ref{eq:scatter step 1})-(\ref{eq:scatter step 4}),
is limited by a Courant condition, i.e., ion should not cross multiple
wavelength cells during a single scattering. This is unfeasible in
practice, as a given change in pitch maps into larger and larger changes
in $\lambda$ as the particle energy increases (see Eqn. {[}\ref{eq:resonant wavelength}{]}).
Noting that the wavelength cells are very small near $\lambda=0$,
and that $a\propto dt^{1/2}$, obeying Courant condition would result
in prohibitely small time step.

Additional limitation to scattering time step comes from an analytic
result of the Fokker-Planck Eqn. (\ref{eq:fokker-planck}), namely
that a time-independent solution must be a function of state variables
$(v,\mu)$ only. In other words, an initially isotropic distribution
in $\mu$ must stay isotropic at all times. The isotropicity condition
forces the time step to be extremely small near $\tilde{\mu}=0$ unless
$a$ and $b$ are modified \citep[see, e.g,][for discussion]{2013ApJS..207...29A}.
Our modification is as follows: if $\left|\tilde{\mu}\right|<\tilde{\mu}_{\mathrm{min}}$,
resonant wavelength is calculated using $\tilde{\mu}_{\mathrm{min}}$
instead of the correct $\tilde{\mu}$ when fetching the wave intensities
above. This will ``freeze'' $a$ to a constant value in the $\left|\tilde{\mu}\right|<\tilde{\mu}_{\mathrm{min}}$
interval, similarly to a technique used in \citet{2008AIPC.1039.....L,2012ApJ...761..104M}.
We use a value $\tilde{\mu}_{\mathrm{min}}=0.025$ here. 

\begin{figure}[!h]
\begin{centering}
\includegraphics[width=1\columnwidth]{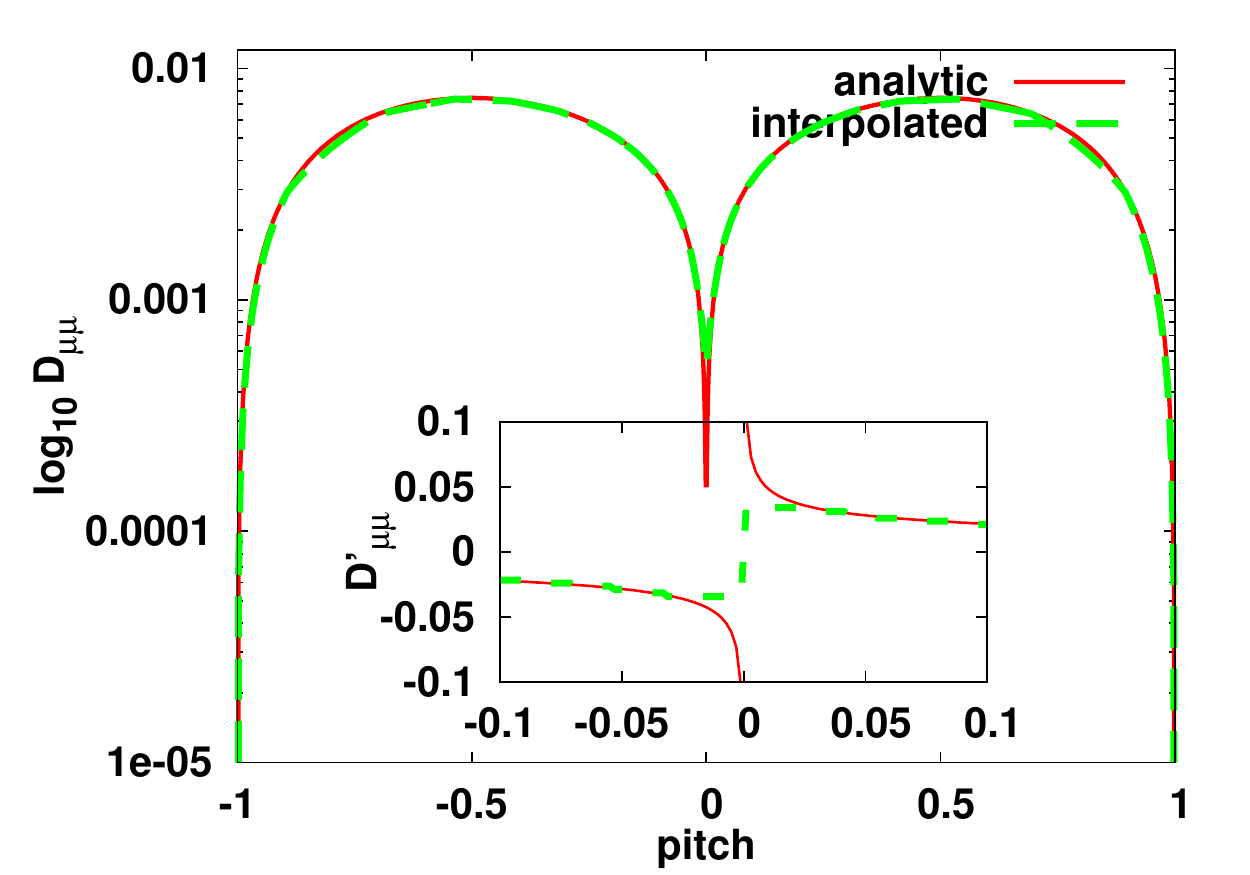}
\par\end{centering}

\caption{Interpolated (dashed, colored green in electronic edition) vs. analytic
(solid, colored red in electronic edition) pitch angle diffusion coefficient
$D_{\mu\mu}$ for wave intensity $I(k)=I_{0}\left(k/k_{0}\right)^{-5/3}$.
Inset shows interpolated vs. analytic $\partial_{\mu}D_{\mu\mu}$
near $\mu=0$, illustrating the effect of our modification. See the
electronic edition of the Journal for a color version of this figure.\label{fig:Interpolated vs analytic Dmumu}}
\end{figure}

Figure \ref{fig:Interpolated vs analytic Dmumu} shows modified drift
and diffusion terms for a power law intensity for a $110$ MeV proton.
Wave intensity is the same as in the simulations discussed in Section
\ref{sec:Results}. The drift term gets very large values near $\mu=0$,
which we effectively remove by ``filling'' the $\tilde{\mu}=0$
resonance gap.

We impose a limitation that the maximum change in pitch is $\Delta\mu_{\mathrm{max}}$
during a single scattering, and substep the scatterer if the simulation
time step $dt_{\mathrm{sim}}$ exceeds the maximum scattering time
step. Scattering time step is evaluated as 

\begin{equation}
dt_{\mathrm{max}}=\mathrm{min}\left(\frac{\Delta\mu_{\mathrm{max}}}{a_{\mathrm{max}}},\frac{\Delta\mu_{\mathrm{max}}^{2}}{b_{\mathrm{max}}^{2}}\right),\label{eq:max scatter time step}
\end{equation}
where $a_{\mathrm{max}}$ and $b_{\mathrm{max}}$ are estimates for
maximum absolute values of the drift and diffusion terms that the
particle can encounter. The number of substeps is $N_{\mathrm{sub}}=dt_{\mathrm{sim}}/dt_{\mathrm{max}}$.
We precalculate the number of required substeps%
\footnote{The relativistically incorrect speed of $1.4\cdot10^{9}$ $\mathrm{m}\:\mathrm{s}^{-1}$
is only used to estimate the number of needed substeps.%
} for speeds $\tilde{V}\in1.4\cdot\left[10^{6},10^{7},10^{8},10^{9}\right]$
$\mathrm{m}\:\mathrm{s}^{-1}$ for all spatial cells. We then pick
a suitable number of substeps for each particle based on the precalculated
values. We keep track of the substeps the scatterer takes, and adjust
$dt_{\mathrm{sim}}$ to keep $N_{\mathrm{sub}}\leqslant10$.

\subsubsection{Injection}

The statistical weight $W$ of a macroparticle is the number of real
particles it represents within some energy and pitch intervals,

\begin{equation}
W=n(\mathbf{r})\Delta x^{3}\int_{U}^{U+\delta U}\int_{\mu}^{\mu+\delta\mu}f(U,\mu)dUd\mu.\label{eq:macroparticle weight}
\end{equation}
where $f(U,\mu)=\left(1/2\right)f_{\kappa}(U)\left[\theta(\mu+1)-\theta(\mu-1)\right]$,
and $\theta$ is the step function.

Macroparticles are injected at a constant rate of $N$ macroparticles
per second. Thus, each time step only $N_{\mathrm{inj}}=N\Delta t$
macroparticles are injected, and the fractional part of $N\Delta t$
is used as a probability to inject an additional macroparticle. The
injection energy interval is divided into $N_{\mathrm{inj}}$ uniformly
spaced bins. Injected macroparticle gets a random energy in its bin,
and a uniformly distributed random pitch $\mu\in\left[-1,+1\right]$.
$W$ is integrated using the bin limits in Eqn. (\ref{eq:macroparticle weight}).

The method described above gives a rather uniform distribution of
particles over the injection energy range. If macroparticle $U$ was
randomized over the whole injection energy range instead of using
the binning method, occasionally several low energy macroparticles
could be generated, which would cause large (of the order of few)
deviations from the correct number density. We also check statistical
weights against spectral wave energies (see Section \ref{sub:Particle Splitters}),
and split injection ions as many times as necessary to keep the simulation
stable.

\subsection{Wave Growth Factors\label{sub:Wave-Growth-Factors}}

A particle subject to scattering $\Delta\tilde{\mu}$ changes its
energy in plasma frame by $\Delta U=mV_{\mathrm{A}}\tilde{V}\Delta\tilde{\mu}$,
which must be removed from waves to conserve the total (particles+waves)
energy. Resonant wavelength changes by $\Delta\lambda_{\mathrm{r}}=\left(2\pi/\Omega\right)\tilde{V}\Delta\tilde{\mu}$.
Thus, the exchange energy per unit wavelength is

\begin{equation}
\Delta u=W\frac{mV_{\mathrm{A}}\Omega}{2\pi}.\label{eq:particle energy change}
\end{equation}
For each wavelength cell the particle fully crosses due to scattering,
the wave energy changes by $\Delta U_{\mathrm{w,}\mathrm{l}}=-\Delta u\Delta\lambda_{\mathrm{l}}$,
where $\Delta\lambda_{\mathrm{l}}$ is the (physical) cell size. Each
spatial cell receives an energy change $\Delta U_{il}=S_{i}\Delta U_{\mathrm{w,}l}$,
i.e., shape factors are only used in $x$ direction. Partial wavelength
cell crossings are accounted for in the same way, but the exact distance
traveled is used instead of the full cell width $\Delta\lambda_{\mathrm{l}}$
when calculating $\Delta U_{\mathrm{w,l}}$.

After wave energy changes have been accumulated a diffusive smoothing
is applied,

\begin{equation}
\frac{\delta\left(\Delta U_{\mathrm{l}}\right)}{\Delta\lambda_{\mathrm{l}}}=-D\left[\frac{\Delta U_{\mathrm{l+1}}}{\Delta\lambda_{\mathrm{l+1}}}-\frac{2\Delta U_{\mathrm{l}}}{\Delta\lambda_{\mathrm{l}}}+\frac{\Delta U_{\mathrm{l-1}}}{\Delta\lambda_{\mathrm{l-1}}}\right]\label{eq:diffusive smoothing}
\end{equation}
with diffusion coefficient $D=0.25$ to smooth out fluctuations due
to numerical noise. Eqn. (\ref{eq:diffusive smoothing}) is conservative,
provided that diffusive fluxes are set to zero at the boundaries of
wavelength mesh. Smoothing is done twice each time step before wave
packet energies are modified.

Finally, the energy changes are applied to wave packets. Each wave
packet changes its energy by $\Delta U_{\mathrm{packet}}=S_{\mathrm{il}}\left(U_{\mathrm{packet}}/U_{\mathrm{w,}\mathrm{il}}\right)\Delta U_{\mathrm{w,}\mathrm{il}}$.
The extra factor in parenthesis distributes the energy change $\Delta U_{\mathrm{w,}\mathrm{il}}$
to all wave packets neighboring that phase-space cell ($U_{\mathrm{w,}\mathrm{il}}=\sum_{\mathrm{n}}S_{\mathrm{il}}^{(n)}U_{\mathrm{packet,}\mathrm{n}}$
is the accumulated sum of $\mathrm{n}$ wave packet energies).

\begin{figure}[!t]
\begin{centering}
\includegraphics[width=1\columnwidth]{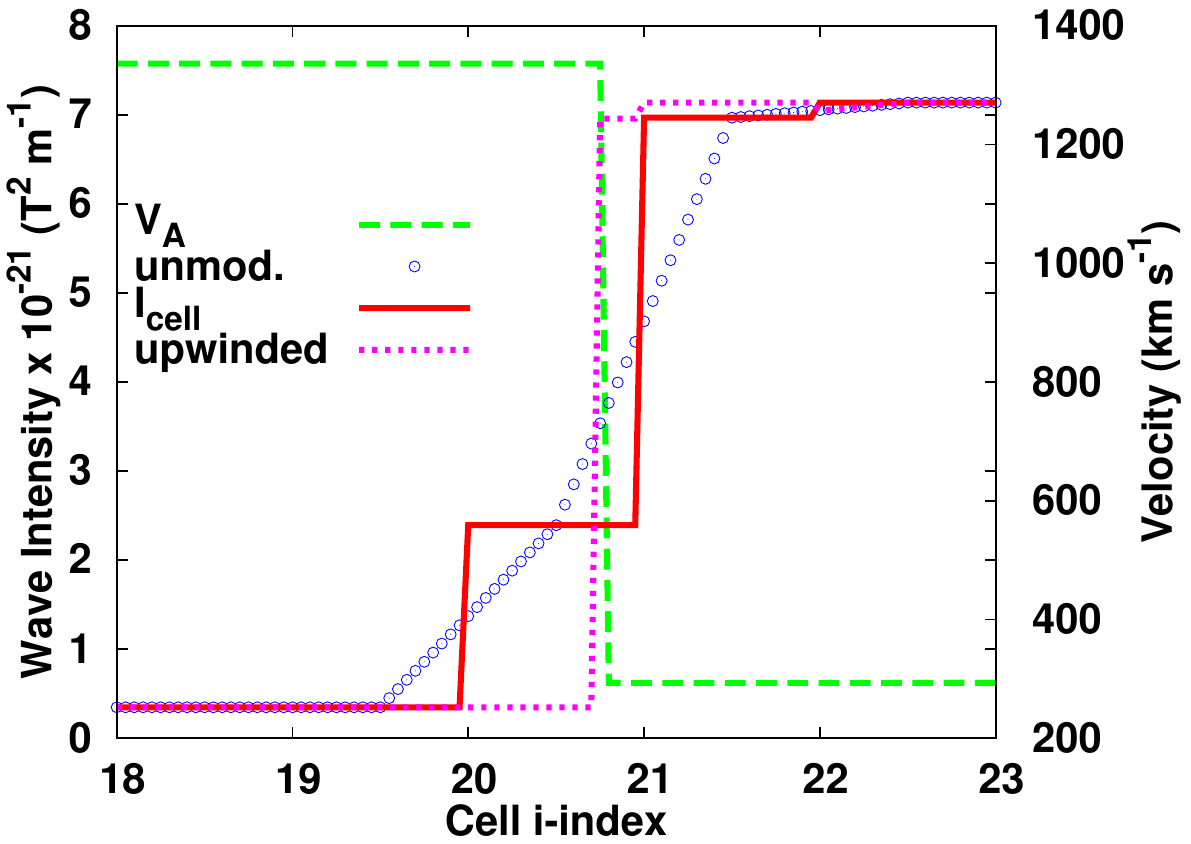}
\par\end{centering}

\caption{Interpolated wave intensity with (dotted, colored purple in electronic
edition) and without (circles) shock sharpening vs. cell $\mathrm{i}$-index.
Shown are also wave intensity in cells (solid, colored red in electronic
edition), and parallel Alfvén speed in shock normal incidence frame
(dashed, colored green in electronic edition), which indicates the
location of the shock front. See the electronic edition of the Journal
for color version of this figure.\label{fig:shock sharpening}}
\end{figure}

\subsection{Particle Splitters\label{sub:Particle Splitters}}

We use two particle splitters in the simulation. The first one stabilizes
the simulation by splitting macroparticles that have too large statistical
weight relative to spectral wave energy. During the scattering process,
a macroparticle can move anywhere within $\lambda\leqslant\left|\lambda_{\mathrm{r,max}}\right|=\left(2\pi/\Omega\right)\tilde{V}$
in $\lambda$ direction. If the relative energy change in any wavelength
cell, given by ratio (see Section \ref{sub:Wave-Growth-Factors})

\begin{equation}
R=\Delta u\Delta\lambda_{\mathrm{l}}/U_{\mathrm{w,min}},\label{eq:energy change ratio}
\end{equation}
where $U_{\mathrm{w,min}}=U_{\mathrm{w}}\left(\left|\lambda\right|\leqslant\left|\lambda_{\mathrm{r,max}}\right|\right)$
is the minimum wave energy the particle can encounter, is too large,
the scattering of a single macroparticle can remove a significant
fraction of the wave energy from that cell or even cause the energy
to become negative. If low wave energy regions are created, the tendency
is for such regions to grow until all wave energy has been removed.

We periodically do a pass over spatial cells and check that macroparticles
do not violate the condition $R\geqslant0.01$. If it is, the macroparticle
is split into 10 new identical particles with statistical weights
$W/10$. This issue mostly concerns macroparticles that have a low
injection energy and a large statistical weight, as low energy ions
are more numerous. Alternatively, we could inject more low energy
particles to keep $W$ small. 

\begin{figure*}[!t]
\begin{centering}
\includegraphics[width=0.33\textwidth]{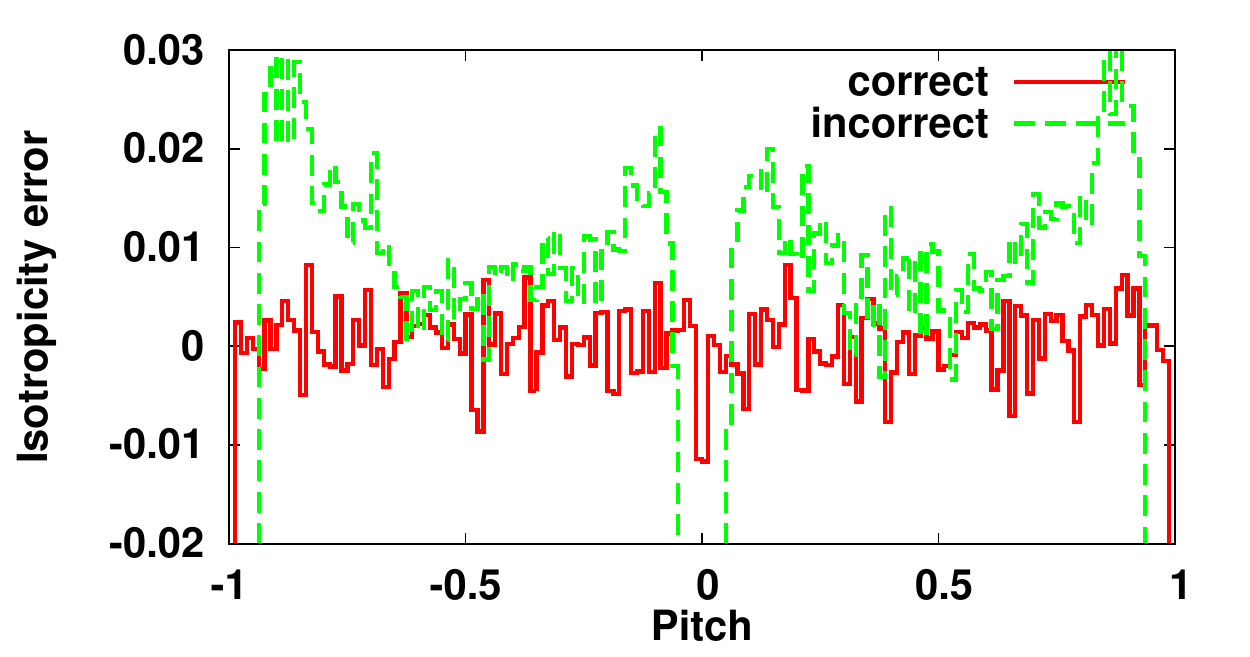}\includegraphics[width=0.33\textwidth]{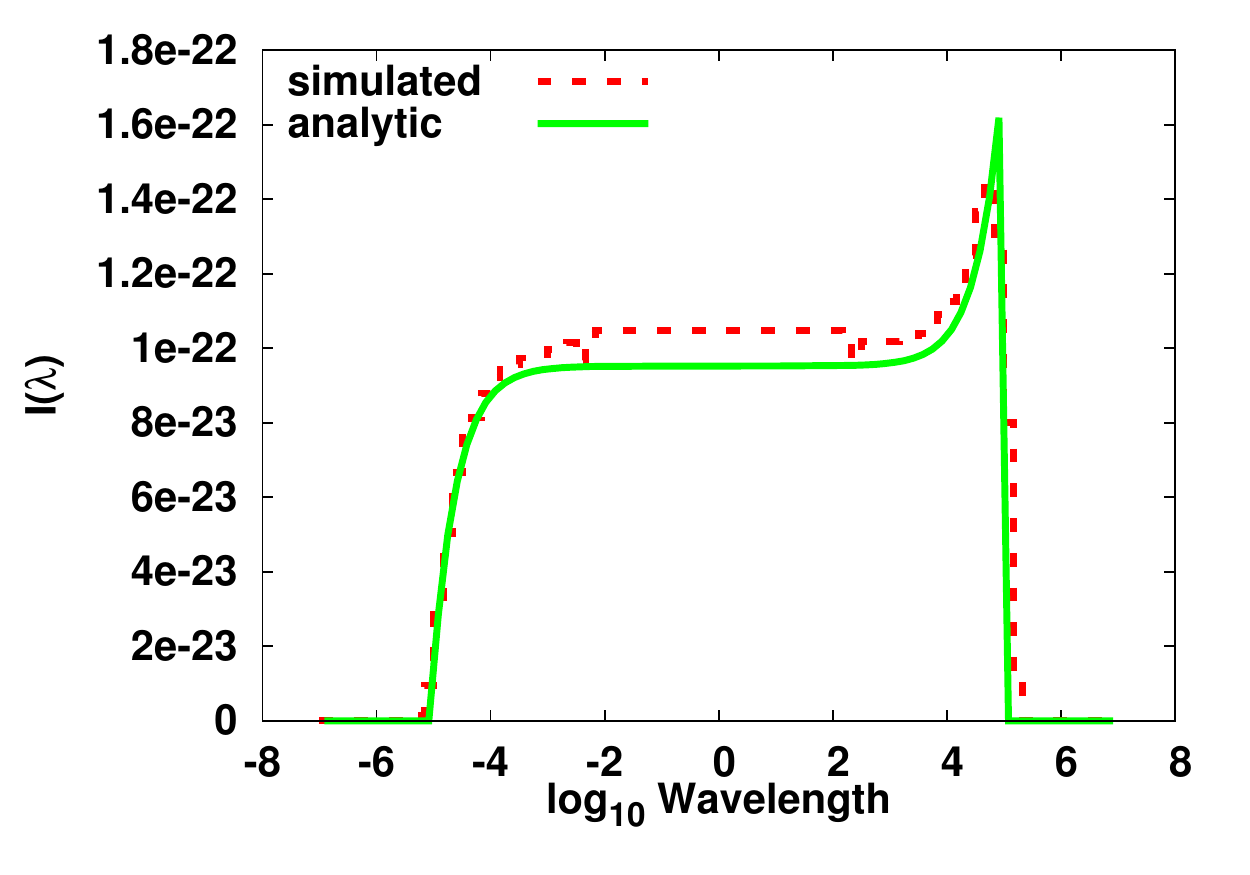}\includegraphics[width=0.33\textwidth]{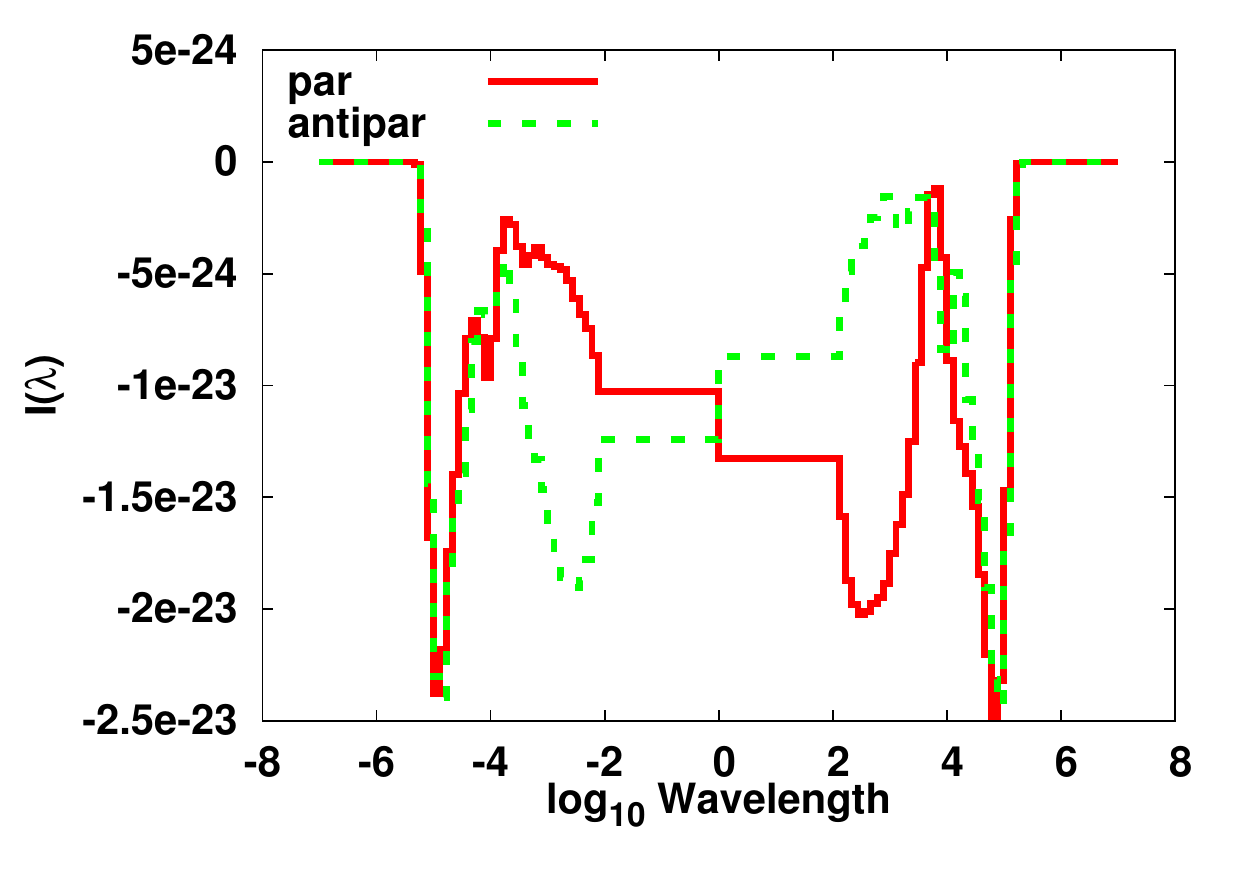}
\par\end{centering}

\caption{Left) Isotropicity error of first-order accurate predictor-corrector
method with correct (solid, colored red in electronic edition) and
incorrect (dashed, colored green in electronic edition) number of
substeps. Middle) Simulated (dashed, colored red in electronic edition)
spectrum of self-generated parallel Alfvén waves by isotropization
of an initial beam distribution of protons vs. analytic result (solid,
colored green in electronic edition). Right) Simulated parallel (solid,
colored red in electronic edition) and antiparallel (dashed, colored
green in electronic edition) self-generated wave spectra for second-order
Fermi acceleration. See the electronic edition of the Journal for
color versions of these figures.\label{fig:Test Cases}}
\end{figure*}

A second particle splitter we use aims to improve resolution at high
energies. We periodically bin the macroparticles based on their current
simulation frame energy, and split particles if a bin contains too
few particles.

\subsection{Shock Sharpening\label{sub:Shock Sharpening}}

Pitch angle diffusion coefficients are calculated using interpolated
wave intensities, while the wave speeds are calculated using analytic
expressions (see Section \ref{sub:Scattering}). Interpolations incresed
shock width to 3-4 spatial cell widths, as illustrated by the unmodified
interpolation curve (symbols) in Figure \ref{fig:shock sharpening}.
Problem this caused was that protons near the shock did not experience
correct scattering center compression ratio. Thus, the first-order
Fermi acceleration ended up being too slow and resulting proton power
law indices were incorrect.

We solve this issue by upwinding (downwinding) particle shape factors
if the particle is near the shock in the upstream (downstream) region.
First, the spatial cell in which the shock front resides is calculated
($\mathrm{i}=20$ cell in Fig. \ref{fig:shock sharpening}). Then,
shape factors are modified so that protons in neighboring cells ($\mathrm{i}\in\left[19-21\right]$
in Fig. \ref{fig:shock sharpening}) do not use the wave intensity
values in the shock cell ($\mathrm{i}=20$ in Fig. \ref{fig:shock sharpening}).
Instead, the values are picked from the neighboring upstream (downstream)
cell if the proton is in the upstream (downstream) region.

Related issue came up with the accumulation of wave energy changes
and interpolations to wave packet positions (see Section \ref{sub:Wave-Growth-Factors}).
Consider protons scattering in the spatial cell immediately downstream
of the shock ($i=21$ in Fig. \ref{fig:shock sharpening}). If the
protons accumulate wave growth factors to upstream side of the shock
($i=19,\,20$ cells in Fig. \ref{fig:shock sharpening}), those growth
factors will be effectively multiplied by the wave transmission coefficient
when the waves cross the shock front, leading to incorrect wave intensities
in the downstream region.

We solved this in a similar manner for the intensity interpolations.
Wave packets in the upstream region are only allowed to accumulate
energy to upstream cells ($i\leqslant20$ cells in Fig. \ref{fig:shock sharpening}),
while wave packets in the downstream region can only accumulate to
downstream cells ($i\geqslant21$ in Fig. \ref{fig:shock sharpening}).
Same condition is also applied when accumulating wave energy changes
from proton positions to phase-space mesh, and when interpolating
accumulated values to wave packet positions. Rest of the wave growth
algorithm in Section \ref{sub:Wave-Growth-Factors} is unchanged.
This solution prevents wave growth from being applied to wrong side
of the shock front, while simultaneously conserving energy within
roundoff errors.

\section{Test Cases\label{sec:Test-Cases}}

\subsection{Isotropicity\label{sub:Isotropicity}}

According to the Fokker-Planck Eqn. (\ref{eq:fokker-planck}), an
initially isotropic particle distribution in pitch must remain isotropic
at all times. This property can be used to estimate the error of the
scattering algorithm.

We test the isotropicity condition by evaluating a master equation
for the scattering process as follows: a simulation is launched with
a single spatial cell and wavelength mesh limits $\pm\lambda_{\mathrm{max}}$.
Wave intensity is taken to be a power law. Wave packets are injected
to the simulation, and their energies are accumulated to the simulation
mesh to form the spectral intensity.

Consider particles having maximum resonant wavelengths $\lambda_{\mathrm{r,max}}=\left(\mathrm{2\pi/\Omega}\right)V=C\cdot\lambda_{\mathrm{max}}$,
where $C\in\left[0,1\right]$ is a constant. Divide the pitch interval
$\mu\in\left[-1,+1\right]$ into $N$ bins of width $\Delta\mu=2/N$,
and form an $N$ by $N$ matrix $A_{\mathrm{ij}}$ for storing the
results. For each column $\mathrm{j}$, scatter $M$ particles with
a random pitch in interval $\mu_{\mathrm{j}}\leqslant\mu\leqslant\mu_{\mathrm{j}}+\Delta\mu$,
and add each particle to row $i$ of the matrix based on the new pitch
$\mu'$ obtained from the scattering algorithm. The quantity $\delta M_{j}=\left(M-\sum_{i}A_{ij}\right)/M$
is the isotropicity error of the scattering algorithm.

Figure \ref{fig:Test Cases}a shows the isotropicity error for 110
MeV protons using $M=100000$ particles and $N=160$ bins. Wave intensity
was the same as that used in DSA simulations 1-3 (see Section \ref{sec:Results}).
Simulation time step was set to 10 seconds, but the adaptive time
step control set the scatterer $dt$ to 0.02849 seconds (351 substeps).
Figure \ref{fig:Test Cases}a also shows corresponding results but
with incorrect (10) number of  substeps. The scattering algorithm,
with substepping enabled, is able to maintain isotropicity extremely
well.

\subsection{Self-Generated Wave Spectra\label{sub:Self-Generated-Wave-Spectrum}}

Alfvén wave spectrum generated by isotropization of a beam distribution
($\tilde{\mu}=1$ in parallel Alfvén wave rest frame) is given by
\citep{2002A&A...393...69S,2013ApJS..207...29A}

\begin{eqnarray}
I(k) & = & 2\pi nV_{\mathrm{A}}\frac{p\Omega}{Vk^{2}}\left(1+\frac{\Omega}{kV}\right),\\
I(\lambda) & = & \frac{\mu_{0}nV_{\mathrm{A}}p}{2\lambda_{\mathrm{r,max}}}\left(1+\frac{\lambda}{\lambda_{\mathrm{r,max}}}\right).\label{eq:self-generated spectrum}
\end{eqnarray}
Eqn. (\ref{eq:self-generated spectrum}) represents a simple triangle
pulse constrained to interval $\left|\lambda\right|\leqslant\lambda_{\mathrm{r,max}}$.
Self-generated wave spectra from simulations are in very good agreement
with the analytic result (see Figure \ref{fig:Test Cases}b). Results
were produced by running the simulation with a single spatial cell
and applying periodic boundary conditions. Total (wave+particle) energy
is conserved within roundoff errors.

Figure \ref{fig:Test Cases}c shows self-generated wave spectra in
a similar setup as above, but with equal initial parallel and antiparallel
Alfvén wave background. Monoenergetic (86.2 keV, $2.26\:\mu\mathrm{T}$
magnetic field, $n=9.085\cdot10^{10}\:\mathrm{m}^{-3}$ plasma number
density) protons had initially an isotropic distribution in pitch
in plasma rest frame. It is not possible to reach distribution that
is isotropic in both wave rest frames simultaneously, because the
waves move to opposite directions. The second-order Fermi process
will, on average, continuously accelerate particles, and damp both
wave modes. Resulting self-generated wave spectra have very pronounced
``spikes'' at long ($\propto\lambda_{\mathrm{r,max}}$) wavelengths,
and are antisymmetric with each others over $\lambda=0$ point.

This setup is extremely sensitive to numerical noise due to finite
number of macroparticles in the spatial cell. Fig. \ref{fig:Test Cases}c
was produced by averaging results from ten identical simulations that
had different random number generator seeds.

\section{Results\label{sec:Results}}

\subsection{Setup}

\begin{table}[t]
\begin{centering}
\begin{tabular}{|c|c|c|c|c|}
\hline 
Run & Scat$.^{\mathrm{a}}$ & $C_{\mathrm{L,R}}^{+\mathrm{b}}$ & Wave gen$.^{\mathrm{c}}$ & AP$^{\mathrm{d}}$\tabularnewline
\hline 
\hline 
1 & I & $3\cdot10^{-3}$ & N & N\tabularnewline
\hline 
2 & A & $3\cdot10^{-3}$ & N & N\tabularnewline
\hline 
3 & A & $3\cdot10^{-3}$ & Y & Y\tabularnewline
\hline 
4 & A & $6\cdot10^{-3}$ & Y & Y\tabularnewline
\hline 
5 & A & $9\cdot10^{-3}$ & Y & Y\tabularnewline
\hline 
\end{tabular}
\par\end{centering}

\caption{Summary of simulation runs presented in this manuscript. a) Scattering
model used: isotropic (I) or anisotropic (A). b) Normalization constant
of ambient parallel Alfvén wave power spectrum. c) Were protons allowed
to self-generate waves (Y for yes, N for no)? d) Were reflected antiparallel
Alfvén waves created at the shock?\label{tab:Summary of simulation parameters}}
\end{table}

DSA simulations were run with parameters suitable for the solar corona
near 10 solar radii ($\mathrm{R}_{\mathrm{sun}}$). Two-dimensional
simulation box was taken to extend from -10 to +10 solar radii in
$x$-direction, and from $-10^{7}$ to $+10^{7}$ m in wavelength
with $\lambda_{\mathrm{0}}=1000$ m. Upstream plasma number density
was set to $n_{0}=9.085\cdot10^{10}$ $\mathrm{m}^{-3}$, velocity
$1500$ $\mathrm{km}\:\mathrm{s}^{-1}$ in shock rest frame, and magnetic
field $2.26\;\mu\mathrm{T}$, roughly corresponding to the values
of coronal model by \citet{2003A&A...400..329M} at $10\;\mathrm{R}_{\mathrm{sun}}$.
In the simulation box the shock is centered at $x=2.45\;\mathrm{R}_{\mathrm{sun}}$.
Shocks in corona are typically strong, having gas compression ratios
above 3.5 near CME nose regions. Here we simply use a constant gas
compression ratio $R_{\mathrm{gas}}=4$. 

\begin{figure}[!h]
\begin{centering}
\includegraphics[width=1\columnwidth]{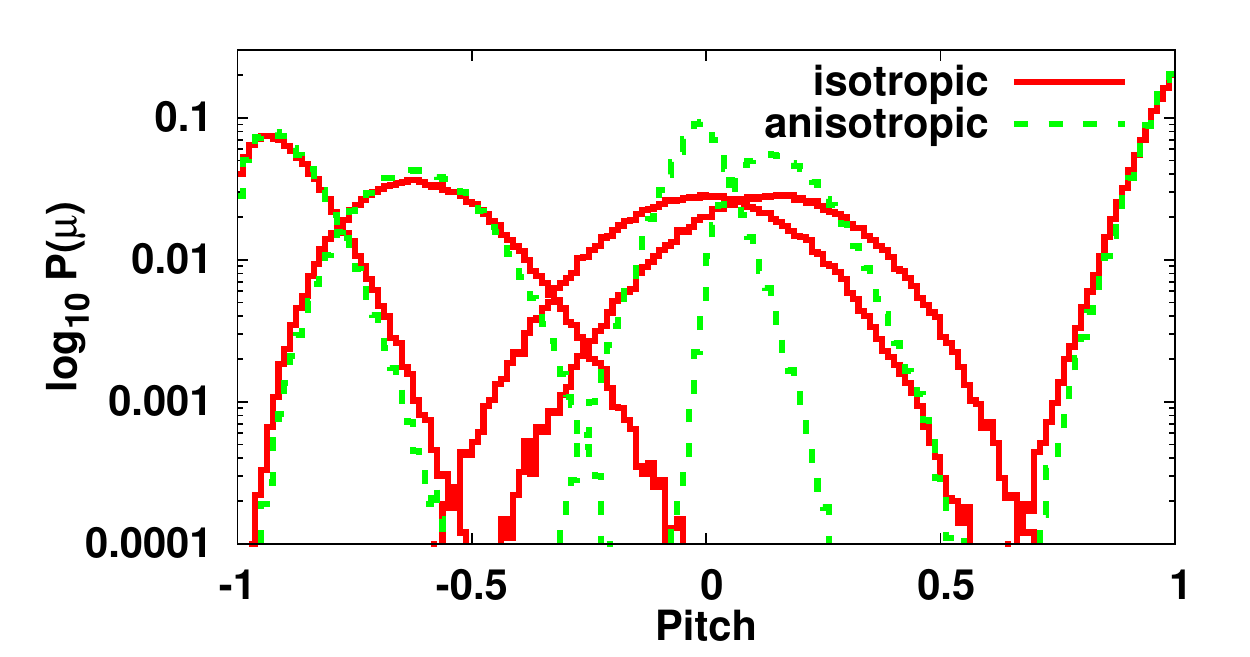}
\par\end{centering}

\caption{Comparison of scattering probability densities between isotropic (solid,
colored red in electronic edition) and anisotropic (dashed, colored
green in electronic edition) scattering operators for 110 MeV protons
with several initial pitch values (not shown) for Kolmogorov wave
spectrum. For initial pitch $\tilde{\mu}=\pm1$ the probability densities
are almost equal, but the differences grow larger when initial pitch
$\tilde{\mu}\rightarrow0$. Isotropic scattering is always more efficient
at scattering particles. See the electronic edition of the Journal
for a color version of this figure.\label{fig:probability densities} }
\end{figure}

Parallel-propagating Alfvén wave packets were injected to $-x$ inflow
boundary at a constant rate. A Kolmogorov wave intensity with $k-$spectral
index of $-5/3$ was used, and both helicities were separately normalized
against ambient magnetic field energy density with a normalization
constant $C_{\mathrm{L,R}}^{+}$ over the wavelength range $0\leqslant\left|\lambda_{\mathrm{max}}\right|\leqslant4\cdot10^{8}$
m. To set up the wave background, the simulation was run with Alfvén
waves only for 40000 seconds. Initially, the parallel mean free path
in runs 1-3, \citep{1970ApJ...162.1049H}

\begin{equation}
d_{\parallel}=\frac{3\, v}{8}\int_{-1}^{+1}\frac{\left(1-\mu^{2}\right)}{D_{\mu\mu}}d\mu=3D_{\parallel}/v,\label{eq:parallel mean free path}
\end{equation}
was $5.72\:\mathrm{R}_{\mathrm{sun}}$ in the upstream region, and
$0.17\:\mathrm{R}_{\mathrm{sun}}$ in the downstream region for 0.16-0.32
MeV protons with the chosen parameters. In Eqn. (\ref{eq:parallel mean free path})
$D_{\parallel}$ is the parallel diffusion coefficient. 

In shock rest frame, parallel wave speed in upstream region in 1336
$\mathrm{km}\:\mathrm{s}^{-1}$, and $293.2$ $\mathrm{km}\:\mathrm{s}^{-1}$
in downstream region, while antiparallel wave speed is 456.8 $\mathrm{km}\:\mathrm{s}^{-1}$
in downstream region. Numerically, the transmission and reflection
coefficients for incident wave intensity at constant $\lambda$, are
$32.2$ and $1.1$, which are equal to those obtained from Eqns. (20-21)
in \citet{1998A&A...331..793V}.

After wave background was set up, monoenergetic (1000 Maxwell-Boltzmann
thermal energy, 86.2 keV) protons were injected directly in front
of the shock. Particles with velocity vectors pointing away from the
shock were rejected. Macroparticle weights were using Eqn. (\ref{eq:kappa energy distrib})
with a kappa-index $\kappa=-3$. We ended up multiplying the injected
particle weights by 50000 in all simulation runs to clearly demonstrate
the effects of the wave generation, as the kappa distribution has
very few suprathermals at high energies. The scaled suprathermal number
density was $\sim0.0088\cdot n_{0}$. Simulations were then run for
8000 seconds with various features turned on or off (see Table \ref{tab:Summary of simulation parameters}).

In order to have a comparison against previous modeling efforts, we
ran the same setup using both anisotropic and isotropic pitch angle
scattering operators. Main difference between the operators is that
in isotropic scattering, the particle pitch is neglegted in resonance
condition, i.e., $D_{\mu\mu}$ is evaluated using $\lambda_{\mathrm{r,max}}$
instead of $\lambda_{\mathrm{r,max}}\left|\tilde{\mu}\right|$. We
chose an isotropic scattering algorithm that has been used in several
previous studies \citep[e.g.,][]{1991SSRv...58..259J,1996SoPh..166..135T}. 

Figure \ref{fig:probability densities} shows probability densities
for isotropic and anisotropic scattering for several initial pitch
values (correspond to peak values of each curve). For each initial
$\tilde{\mu}$, the vertical axis gives the probability of scattering
to that $\tilde{\mu}$ over the used time step. Isotropic scattering
is more efficient at diffusing particles across the $\tilde{\mu}=0$
resonance gap.

\subsection{Energy Spectra}

\begin{figure*}
\begin{centering}
\includegraphics[height=5.5cm]{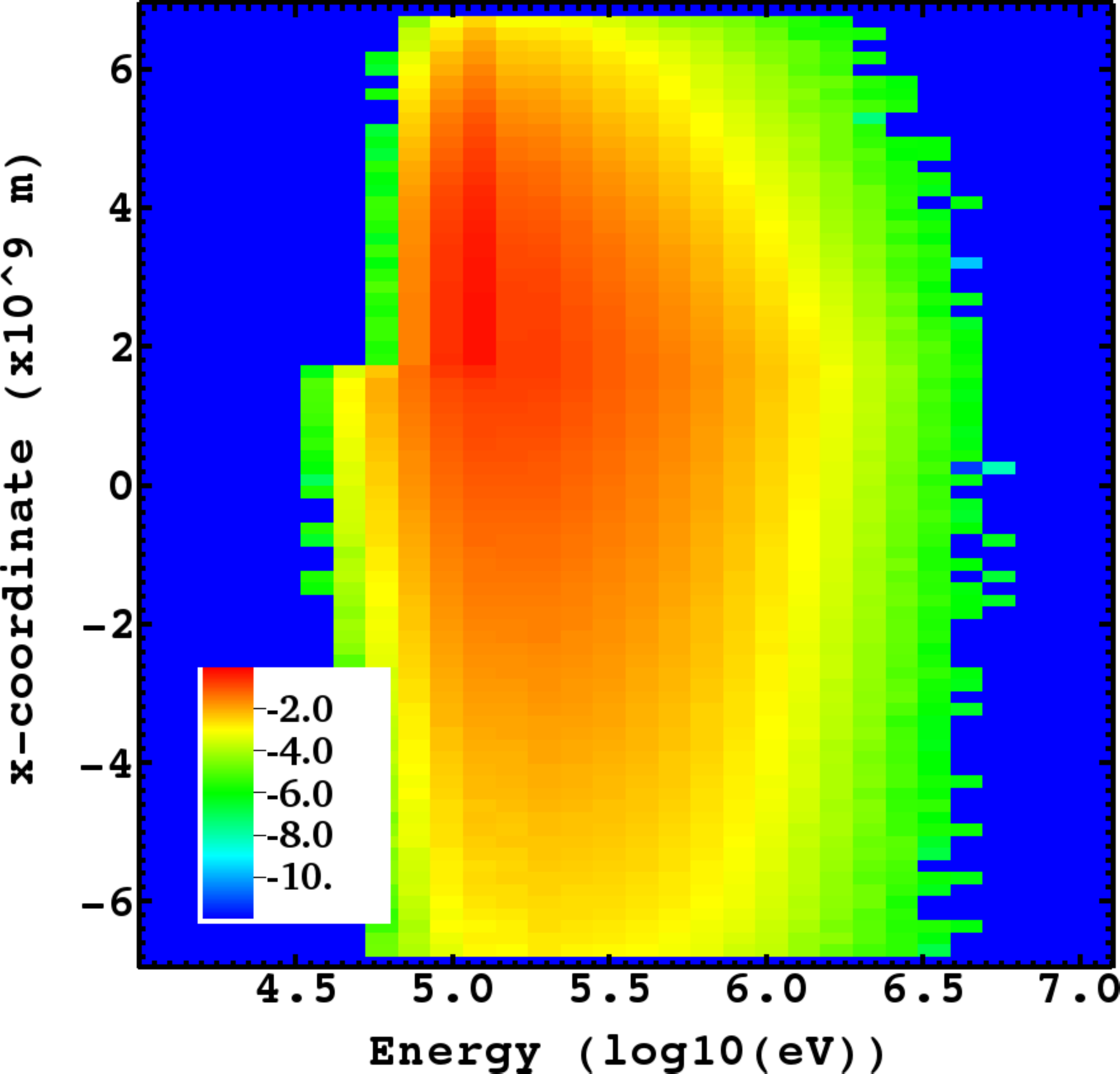}
\includegraphics[height=5.5cm]{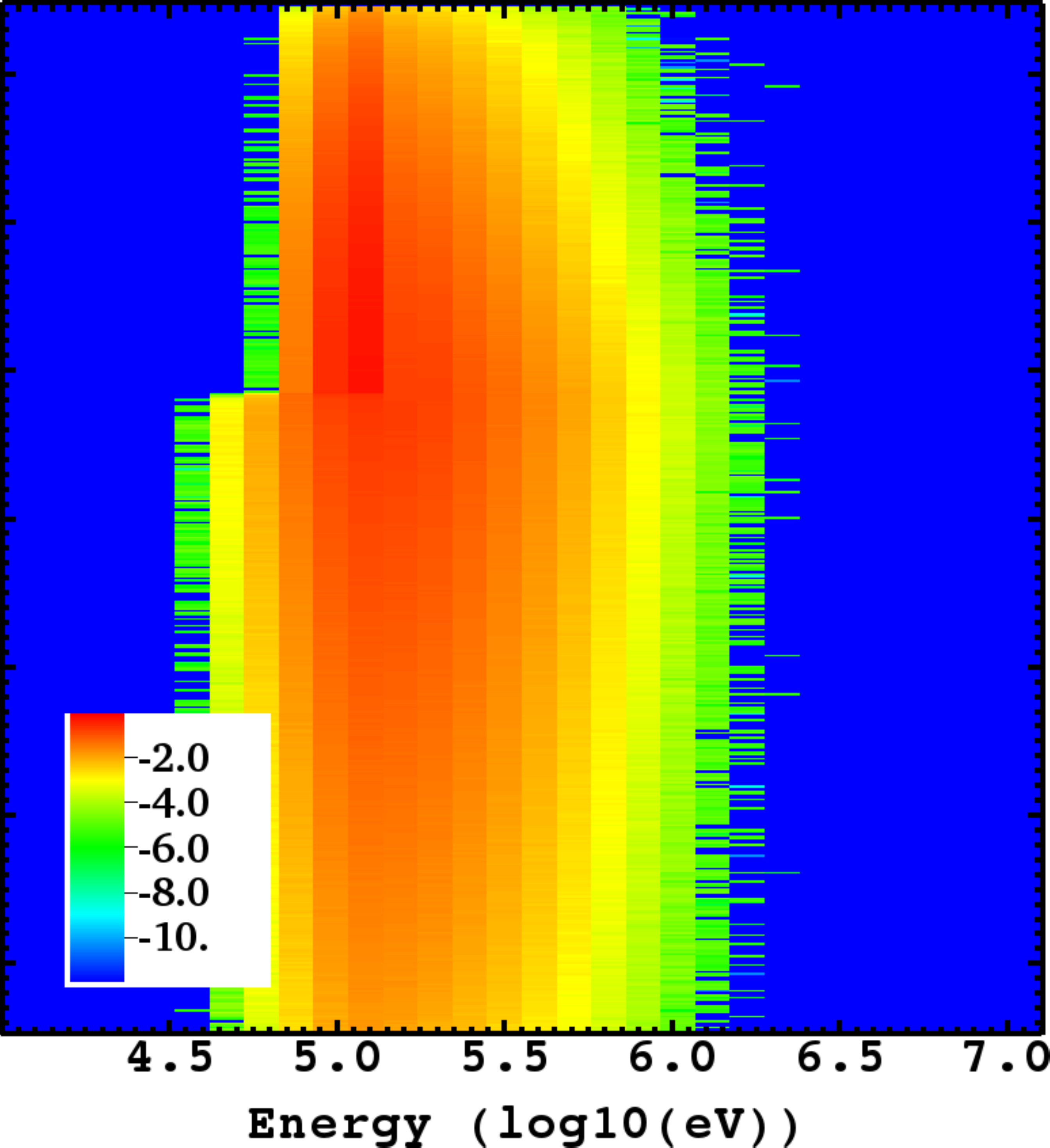}
\includegraphics[height=5.5cm]{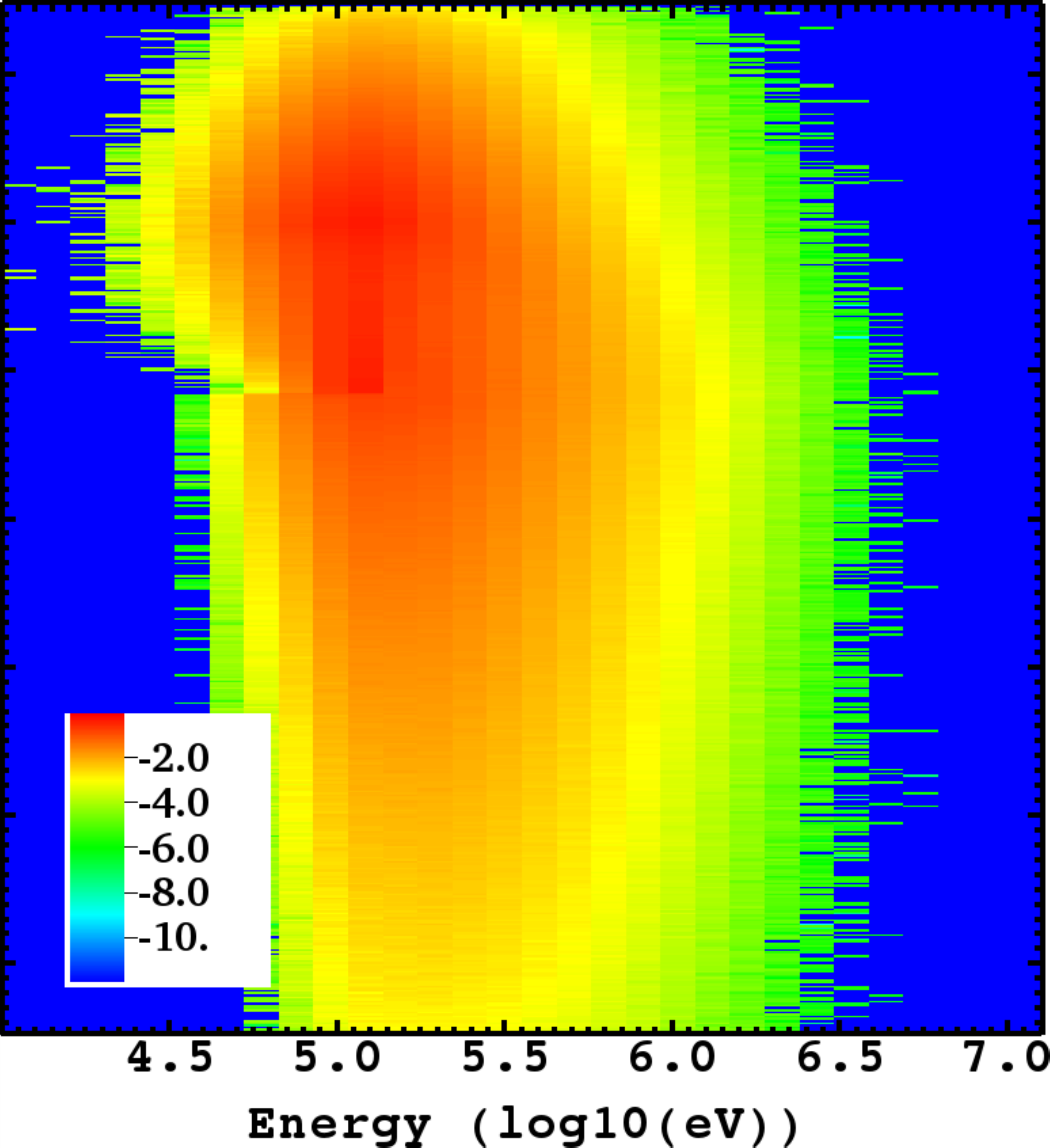}
\par\end{centering}

\caption{Left) Proton energy spectra ($1/\mathrm{m}^{3}\mathrm{eV}$) vs. $x$-coordinate
at the end of simulation run 1 for isotropic scattering and parallel
propagating Alfvén waves only. Middle) Same as left panel, but for
run 2 where anisotropic scattering was used. Isotropic scattering
is more effective at diffusing ions across the resonance gap, and
thus overestimates the rate of diffusive shock acceleration. Right)
Same as middle panel, but for run 4 where reflected antiparallel Alfvén
waves were created and wave self-generation was turned on. The shock
is located near $x=1.71\cdot10^{9}$ m in all panels. See the electronic
edition of the Journal for a color version of this figure.\label{fig:comparison energy spectra}}
\end{figure*}

Figure \ref{fig:comparison energy spectra} shows proton energy spectra
vs. position at the end of simulation runs 1,2, and 4. Left and middle
panels show energy spectra from simulation run where isotropic (run
1) and anisotropic (run 2) scattering operator was used. As expected,
isotropic scattering is more effective in trapping protons near the
shock front, and the maximum energies end up being higher than with
anisotropic scattering. Ultimately, the maximum proton energies are
limited by the size of the simulation box, i.e., the $\pm x$ walls
are free escape boundaries.

The right panel of Fig. \ref{fig:comparison energy spectra} shows
energy spectra from run 4, where antiparallel wave creation at the
shock was turned on, i.e., both first and second-order Fermi acceleration
are operating in this run. Most notable difference to run 2 is the
flattening of energy spectra below $\sim100$ keV due to second-order
Fermi acceleration in the downstream region.

Analytical solutions to similar scenarios than in our simulations
exist, although typically some simplifying assumptions are needed,
for example, for diffusion coefficients, to obtain a solution in closed
form. According to \citet{1983ApJ...270..319W}, if spatial diffusion
coefficients are independent of momentum, and the momentum diffusion
coefficient related to second-order Fermi acceleration decreases exponentially
from the shock in the downstream region, a steady-state momentum distribution
in the upstream region is given by

\begin{equation}
F_{0}(x,p)=F_{0}(0,p)\exp(V_{\mathrm{w,1}}x/D_{\parallel}),\label{eq:Webb upstream solution}
\end{equation}
where $F_{0}(0,p)$ is the isotropic part of proton distribution at
the shock. Thus, distribution at constant momentum should decrease
exponentially with increasing distance $\Delta x$ to the shock with
an e-folding distance

\begin{equation}
d=\frac{V_{\mathrm{w,1}}\Delta x}{D_{\parallel}}=\frac{3V_{\mathrm{w,1}}\Delta x}{v\, d_{\parallel}}.\label{eq:exponential decrease}
\end{equation}

\begin{figure}[!h]
\begin{centering}
\includegraphics[width=1\columnwidth]{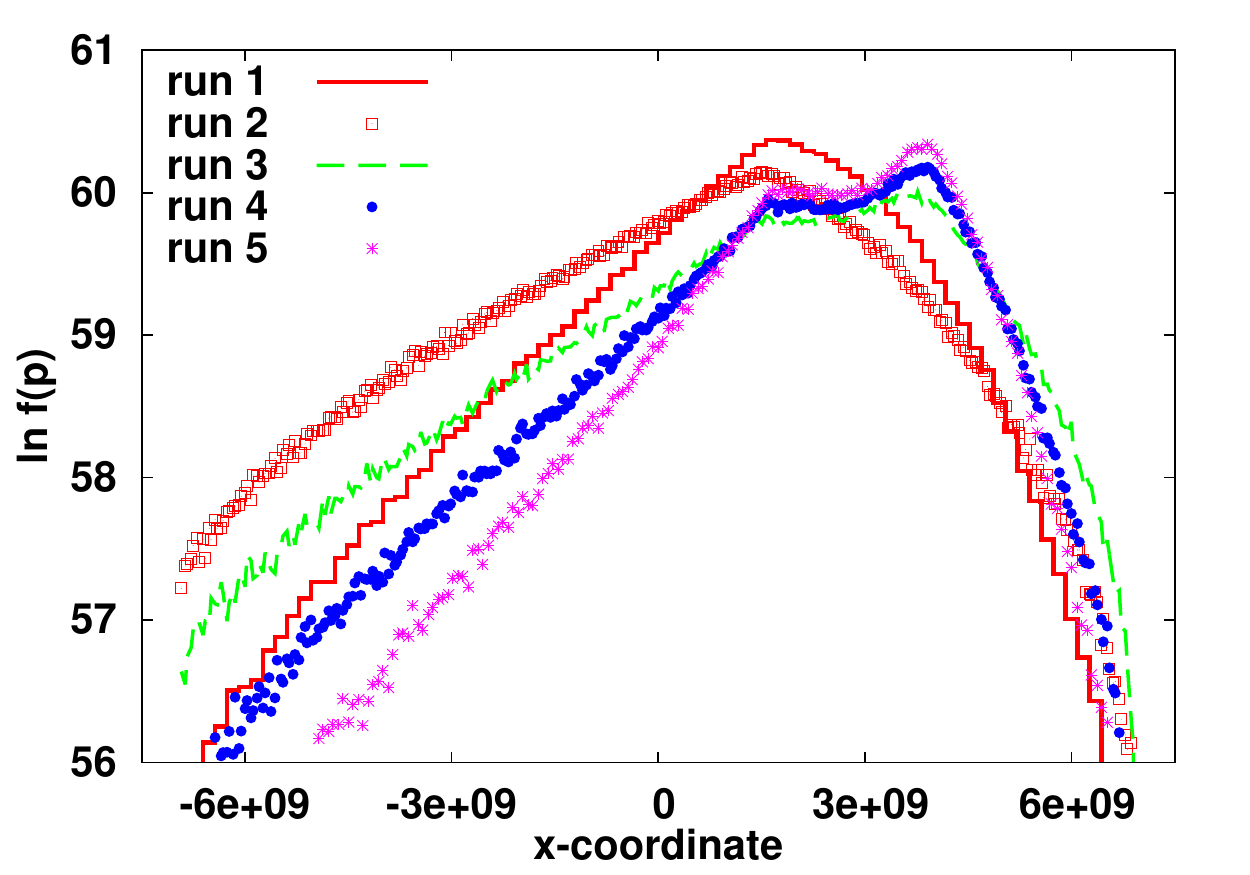}
\par\end{centering}

\caption{Proton momentum distribution at $\sim240$ keV vs. $x$-coordinate
at the end of simulation runs 1-5. In all cases the proton spectra
decrease exponentially with increasing distance to the shock until
boundary effects start to dominate. See the electronic edition of
the Journal for a color version of this figure.\label{fig:Proton spatial momentum spectra}}
\end{figure}

Figure \ref{fig:Proton spatial momentum spectra} shows momentum spectra
at $\sim240$ keV at the end of our simulation runs vs. distance to
the shock. Simulated spectra are in good agreement with the exponential
decrease predicted by Eqn. (\ref{eq:Webb upstream solution}) until
boundary effects start to dominate, i.e., particles hitting $-x$
boundary get removed from the simulation. By using values from simulation
run 3, $\Delta\ln F(p)\approx5.3$, $\Delta x\approx1.1\cdot10^{10}$
m, $V_{\mathrm{w,1}}=1336$ $\mathrm{km}\:\mathrm{s^{-1}}$, and $v=6800$
$\mathrm{km}\:\mathrm{s^{-1}}$ for 240 keV protons, the mean free
path according to Eqn. (\ref{eq:exponential decrease}) should be
$d_{\parallel}\approx1.7\: R_{\mathrm{sun}}$, which is lower than
the actual $5.72\: R_{\mathrm{sun}}$ mean free path in our simulations.

Enhanced distributions (``bumps'') in the downstream region near
$x\sim4\cdot10^{9}$ m in runs 3-5 (see Fig.\ref{fig:Proton spatial momentum spectra}),
are due to limited simulation time. At downstream parallel wave speed
of 293.2 $\mathrm{km}\:\mathrm{s}^{-1}$, waves self-generated immediately
after proton injection have moved to $x\approx4.06\cdot10^{9}$ m
during 8000 s (shock is located at $x\approx1.71\cdot10^{9}$ m).

There are many factors that may contribute to the difference between
the analytic result and order-of-magnitude estimate above. For example,
wave generation somewhat decreases mean free paths near the shock,
$\Delta\ln F_{0}(p)$ above was calculated using total proton distribution
instead of the isotropic part, assumptions made by \citet{1983ApJ...270..319W}
do not apply for our model, and the distance (in mean free paths)
between the shock and free escape boundary depends on proton momentum.
Considering that, the simulations are in good agreement with the analytic
result.

\begin{figure}[!t]
\begin{centering}
\includegraphics[width=1\columnwidth]{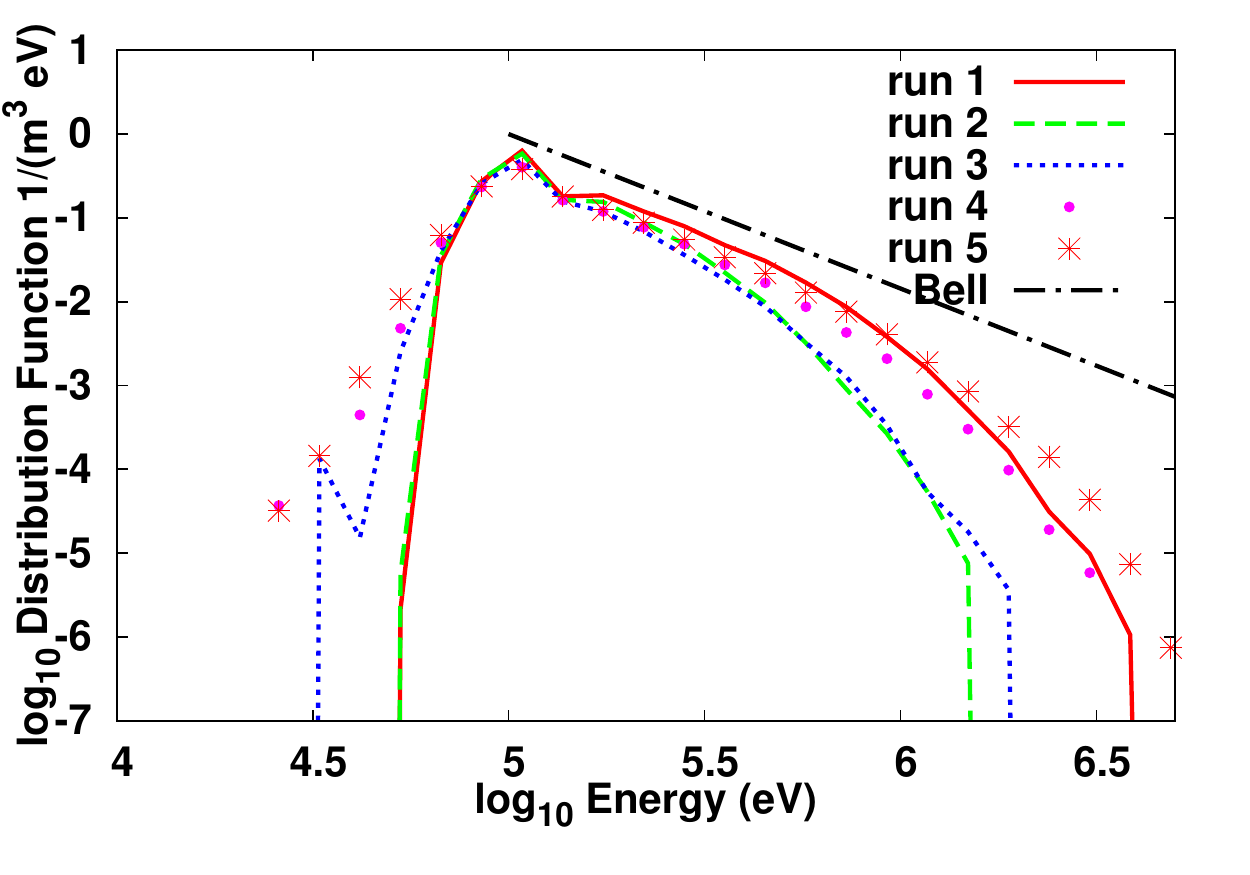}
\par\end{centering}

\caption{Proton energy spectra in the downstream region at $x=2\cdot10^{9}$
m from simulation runs 1-5 vs. analytic prediction (dash-dotted, colored
black in electronic edition). See the electronic edition of the Journal
for a color version of this figure. \label{fig:downstream energy spectra}}
\end{figure}

Well-known result of first-order Fermi acceleration is that the particle
spectrum in the dowstream region is a power law \citep{1977ICRC...11..132A,1977DoSSR.234.1306K,1978MNRAS.182..147B,1978ApJ...221L..29B}

\begin{equation}
f(U)\propto\left(U/U_{0}\right)^{-q},\label{eq:power law index}
\end{equation}
where the spectral index $q=(r+2)/(r-1)=-1.84$ depends only on the
scattering center compression ratio $r=V_{\mathrm{w,1}}/V_{\mathrm{w,2}}=4.56$.
If the time available for acceleration is limited or there are free
escape boundaries, the energy spectra typically show an exponential
cutoff at some energy. Figure \ref{fig:downstream energy spectra}
shows proton energy spectra immediately downstream of the shock at
$x=2\cdot10^{9}$ m from runs 1-5. Spectrum from the isotropic run
1 is in good agreement with the analytic prediction at lower energies
until it starts to soften above $\sim1$ MeV, indicating that protons
at these energies are escaping the acceleration region. Anisotropic
runs need over twice as large ambient wave intensity to reach the
same acceleration efficiency.

\subsection{Self-generated Waves}

\begin{figure}[!t]
\begin{centering}
\includegraphics[width=1\columnwidth]{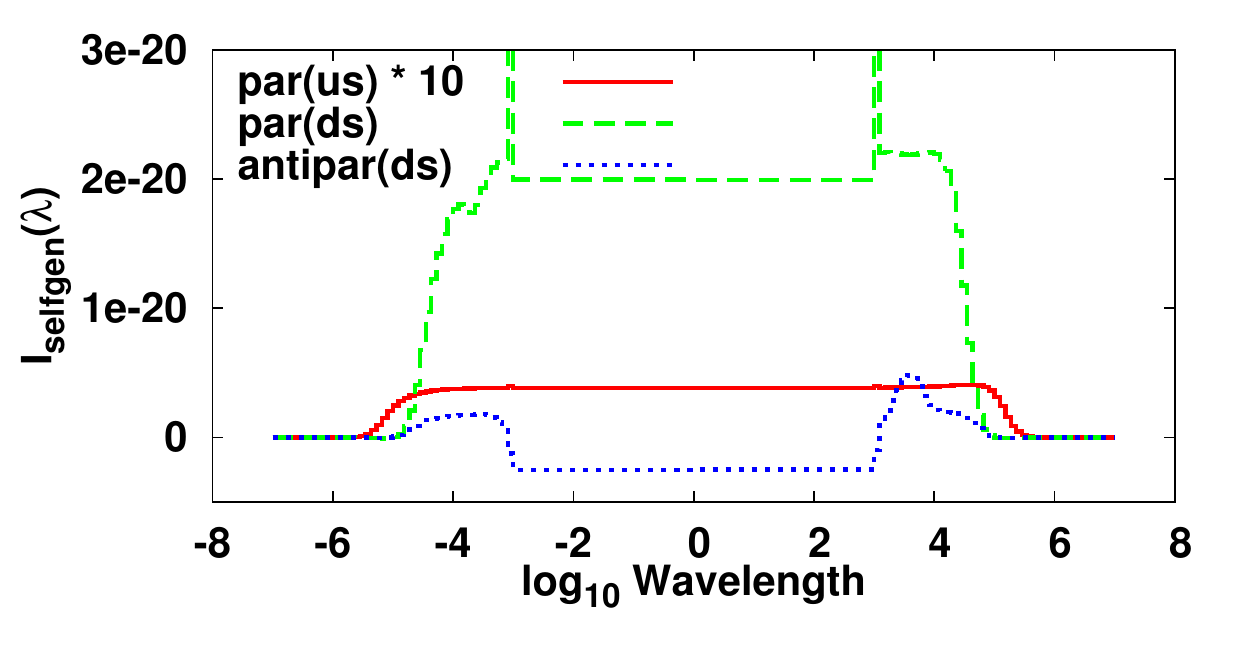}
\par\end{centering}

\caption{Self-generated wave spectra at the end of simulation run 4. Shown
are parallel Alfvén spectra immediately upstream ($x=1.5\cdot10^{9}$,
solid, colored red in electronic edition) and downstream ($x=2\cdot10^{9}$
m, dashed, colored green in electronic edition) of the shock, and
antiparallel Alfvén spectrum downstream of the shock (dotted, colored
blue in electronic edition). The spikes in upstream spectrum at $\lambda=\pm10^{3}$
m are due to changing cell sizes in wavelength mesh. See the electronic
edition of the Journal for a color version of this figure.\label{fig:Self-generated wave spectra}}
\end{figure}

\begin{figure}[!t]
\begin{centering}
\includegraphics[width=1\columnwidth]{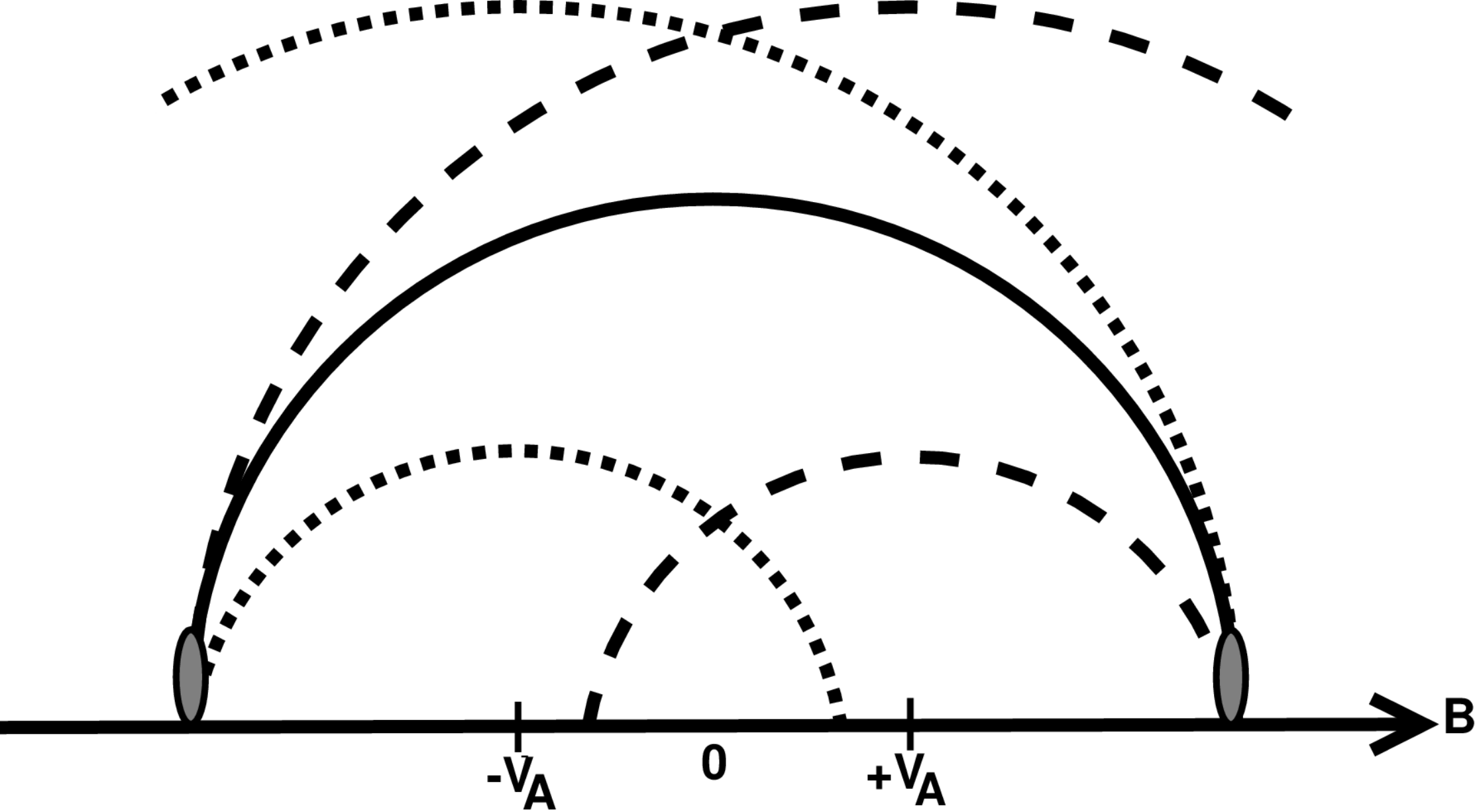}
\par\end{centering}

\caption{Scattering circles for isotropization of initial beam $\tilde{\mu}=\pm1$
distribution (gray ovals). In parallel Alfvén rest frame ($V=+V_{\mathrm{A}}$)
scattering is elastic, and ions move along circles (dashed line).
If the circle in wave rest frame has a smaller (larger) radius than
in plasma rest frame (solid line), ions lose (gain) energy and waves
are amplified (damped). Corresponding results are shown for antiparallel
waves (dotted lines).\label{fig:scattering circles}}
\end{figure}

Figure \ref{fig:Self-generated wave spectra} (solid line) shows the
self-generated parallel and antiparallel wave spectra at the end of
simulation run 4. The characteristic triangle pulse (Sec. \ref{sub:Self-Generated-Wave-Spectrum})
is clearly visible in all spectra. Note the lack of resolution below
$\left|\lambda\right|\leqslant\lambda_{0}=1$ km wavelengths. Various
spectral features in Fig. \ref{fig:Self-generated wave spectra} can
be understood by considering the ``scattering circles'' shown in
Fig. \ref{fig:scattering circles}, which illustrates when wave decay
and amplification occur for an initially beam distribution.

Injected particles initially have an anisotropic $\tilde{\mu}<0$
pitch distribution immediately to the downstream of the shock, thus
isotropization of the transmitted particles damps parallel R-helicity
Alfvén waves, meaning that the peak of self-generated wave spectrum
is located on negative (R-helicity) wavelengths (see Fig. \ref{fig:Test Cases}a).
In Fig. \ref{fig:scattering circles} this roughly corresponds to
the gray oval on left. 

A fraction of injected particles scatter to $\tilde{\mu}>0$ half-plane
and diffuse back to the upstream region, where the isotropization
now amplifies parallel waves (gray oval on right in Fig. \ref{fig:scattering circles}),
and a triangular pulse is created (solid curve in Fig.\ref{fig:Self-generated wave spectra}).
When the amplified waves transmit through the shock, intensity is
amplified, and the width of the pulse reduces by a factor of $\lambda^{\mathrm{T}}/\lambda^{\mathrm{I}}$
(dashed curve in Fig. \ref{fig:Self-generated wave spectra}). Incident
self-generated waves also create antiparallel waves with same helicities
and spectral shape in the downstream region, but at slightly longer
wavelengths (Fig. \ref{fig:Self-generated wave spectra}, dotted curve). 

For antiparallel Alfvén waves in the downstream region the situation
is more complex, as there are several effects operating simultaneously.
First, isotropization of transmitted ions amplifies R-helicity waves
(opposite effect to parallel waves, see Fig. \ref{fig:scattering circles}).
In our simulations this occurs within $\sim5\cdot10^{8}$ m to the
shock. Further into the downstream region, antiparallel waves are
damped in second-order Fermi acceleration. 

\begin{figure}[!h]
\begin{centering}
\includegraphics[width=1\columnwidth]{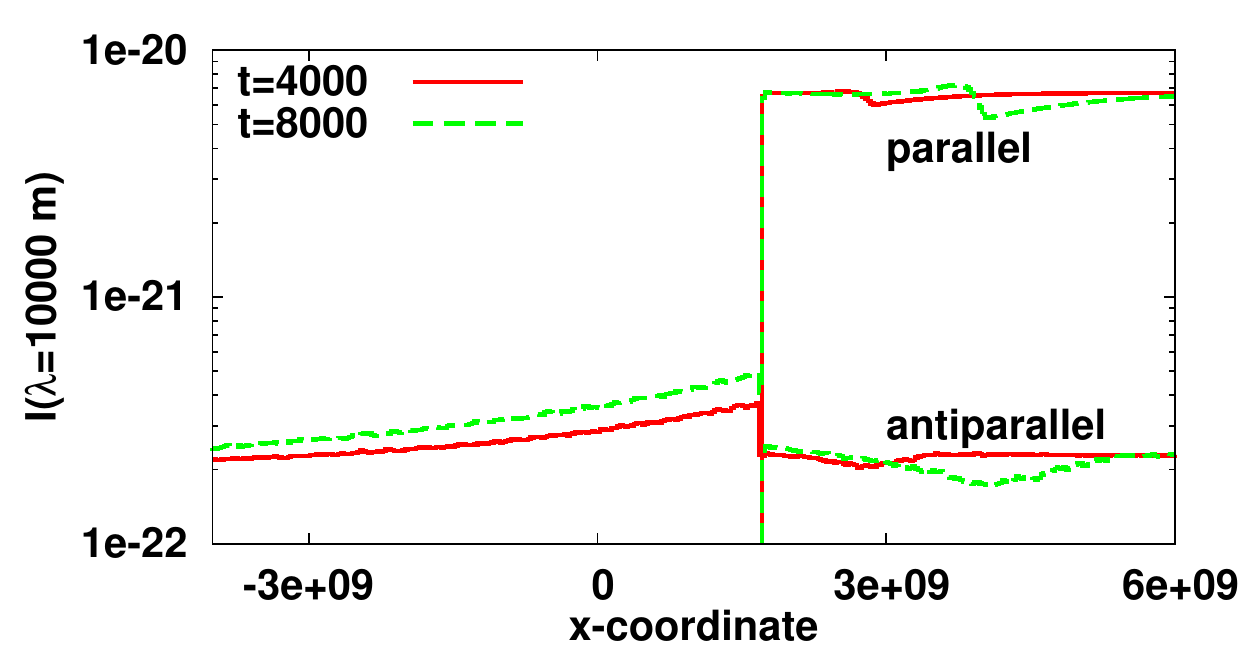}
\par\end{centering}

\caption{Parallel Alfvén wave intensities at $\lambda=100$ km vs. x-coordinate
4000 (solid, colored red in electronic edition) and 8000 (solid, colored
green in electronic edition) seconds after proton injection started
from simulation run 3 where both wave modes were included. Shown are
also antiparallel wave intensities at same values of time (dashed).
See the electronic edition of the Journal for a color version of this
figure. \label{fig:foreshock structure} }
\end{figure}

Figure \ref{fig:foreshock structure} shows parallel wave intensity
at $\log\lambda=5$ vs. $x$-coordinate at 4000 and 8000 seconds into
the simulation from run 3. In the upstream region we see that the
foreshock extends several solar radii from the shock front, and becomes
more pronounced as protons generate more waves. In downstream region
parallel wave intensity stays almost flat, while antiparallel waves
are somewhat damped as discussed above. 

\section{Summary\label{sec:Conclusions}}

We have presented a new numerical model for diffusive shock acceleration
of ions. The described model includes propagation of parallel and
antiparallel slab mode Alfvénic turbulence, transmission of waves
through the shock, and self-consistent wave amplification due to the
accelerating ions. The wave-particle interaction is through gyro resonance.
Waves are treated in a similar fashion as energetic particles. The
new model uses the full form of quasilinear pitch angle scattering
resonance condition, which is a more realistic description than the
simpler isotropic scattering approach used in previous studies, and
is in good agreement with analytic solutions of diffusive shock acceleration.

We have applied the new model to acceleration of protons at parallel
shocks using parameters suitable for solar corona near ten solar radii.
We find that there is a significant difference in efficiency of first-order
Fermi acceleration between isotropic and anisotropic scattering. Isotropic
scattering predicts much higher acceleration rate as compared with
the more realistic anisotropic scattering due to shorter particle
mean free paths. We find that anisotropic scattering requires over
twice as large ambient turbulence in order to reach the same acceleration
efficiency. With the parameter range considered in this study, protons
are accelerated to a few MeV energies. In the future we plan to extend
the model to a more general geometry.

Although the number of suprathermal protons at ten solar radii seems
to be too small for notable wave generation, the resulting wave intensities
show clear deviations from a power law shape near maximum resonant
wavelengths of 30-300 km, especially in the downstream region. These
waves are created by energetic protons that transmitted through the
shock from the downstream to upstream region, where the isotropization
of streaming protons most efficiently amplifies Alfvén waves with
wave vectors pointing away from the shock front. The amplified upstream
waves are transmitted through the shock front, where the compression
of wave magnetic field increases the intensity by a factor of $\sim32$.
Jump in wave speed at the shock causes reflected waves to be created
in the downstream region, that have the same spectral structure as
the incident upstream waves but wave vectors pointing to the opposite
direction.

Wave reflection at the shock enables second-order Fermi acceleration,
and decreases proton mean free paths in the downstream region, somewhat
increasing the efficiency of first-order Fermi acceleration. Second-order
Fermi acceleration, according to our simulations, is generally too
weak process to have a notable effect on the high energy ions. However,
the effect on low energy ($\leqslant100$ keV) ions is significant,
causing considerable flattening of energy spectra. Additionally, we
find that waves in the downstream region are very efficiently damped
by particles, which may be an important mechanism for heating suprathermal
protons. A detailed study of this process is, however, beyond the
scope of this manuscript.

Corsair is a parallel simulation platform, suitable for domain-decomposed
mesh and particle-mesh simulations. Corsair, and the Corsair/SEP shock
acceleration model, are available under GNU general public license,
and can be obtained by contacting AS. Simulations presented here were
carried out on Finnish Meteorological Institute's Cray XT5c supercomputer
using 100 CPU cores.

The work of AS was funded by Academy of Finland grant 251797. The
work at University of Alabama in Huntsville is supported by NSF ATM-0847719.

 \newcommand{\asr}{Adv. Space. Res.}  \newcommand{\jfm}{J. Fluid. Mech.} 

\bibliographystyle{plainnat}
\bibliography{bibliography,/home/sandroos/Dropbox/Publications/SEP_Shock_Sims/bibliography}

\begin{thebibliography}{38}
\providecommand{\natexlab}[1]{#1}
\providecommand{\url}[1]{\texttt{#1}}
\expandafter\ifx\csname urlstyle\endcsname\relax
  \providecommand{\doi}[1]{doi: #1}\else
  \providecommand{\doi}{doi: \begingroup \urlstyle{rm}\Url}\fi

\bibitem[{Afanasiev} and {Vainio}(2013)]{2013ApJS..207...29A}
A.~{Afanasiev} and R.~{Vainio}.
\newblock {Monte Carlo Simulation Model of Energetic Proton Transport through
  Self-generated Alfv{\'e}n Waves}.
\newblock \emph{\apjs}, 207:\penalty0 29, August 2013.

\bibitem[{Axford} et~al.(1977){Axford}, {Leer}, and
  {Skadron}]{1977ICRC...11..132A}
W.~I. {Axford}, E.~{Leer}, and G.~{Skadron}.
\newblock {The acceleration of cosmic rays by shock waves}.
\newblock \emph{Proc. 15th Internat. Cosmic Ray Conf. (Plovdiv)}, 11:\penalty0
  132--137, 1977.

\bibitem[{Battarbee} et~al.(2011){Battarbee}, {Laitinen}, and
  {Vainio}]{2011A&A...535A..34B}
M.~{Battarbee}, T.~{Laitinen}, and R.~{Vainio}.
\newblock {Heavy-ion acceleration and self-generated waves in coronal shocks}.
\newblock \emph{\aap}, 535:\penalty0 A34, November 2011.

\bibitem[{Bell}(1978)]{1978MNRAS.182..147B}
A.~R. {Bell}.
\newblock {The acceleration of cosmic rays in shock fronts. I}.
\newblock \emph{\mnras}, 182:\penalty0 147--156, January 1978.

\bibitem[{Birdsall} and {Langdon}(1985)]{1985PlasmaPhysBirdsall}
C.~K. {Birdsall} and A.~A. {Langdon}.
\newblock {Plasma physics via computer simulation}.
\newblock \emph{New York: McGraw-Hill}, 1985.

\bibitem[{Blandford} and {Ostriker}(1978)]{1978ApJ...221L..29B}
R.~D. {Blandford} and J.~P. {Ostriker}.
\newblock {Particle acceleration by astrophysical shocks}.
\newblock \emph{\apjl}, 221:\penalty0 L29--L32, April 1978.

\bibitem[{Campeanu} and {Schlickeiser}(1992)]{1992A&A...263..413C}
A.~{Campeanu} and R.~{Schlickeiser}.
\newblock {Alfven wave transmission and stochastic particle acceleration at
  parallel astrophysical shock waves}.
\newblock \emph{\aap}, 263:\penalty0 413--422, September 1992.

\bibitem[{Decker}(1988)]{1988SSRv...48..195D}
R.~B. {Decker}.
\newblock {Computer modeling of test particle acceleration at oblique shocks}.
\newblock \emph{\ssr}, 48:\penalty0 195--262, 1988.

\bibitem[{Ellison} et~al.(1990){Ellison}, {Reynolds}, and
  {Jones}]{1990ApJ...360..702E}
D.~C. {Ellison}, S.~P. {Reynolds}, and F.~C. {Jones}.
\newblock {First-order Fermi particle acceleration by relativistic shocks}.
\newblock \emph{\apj}, 360:\penalty0 702--714, September 1990.

\bibitem[{Giacalone}(2005)]{2005ApJ...624..765G}
J.~{Giacalone}.
\newblock {Particle Acceleration at Shocks Moving through an Irregular Magnetic
  Field}.
\newblock \emph{\apj}, 624:\penalty0 765--772, May 2005.

\bibitem[{Hasselmann} and {Wibberenz}(1970)]{1970ApJ...162.1049H}
K.~{Hasselmann} and G.~{Wibberenz}.
\newblock {A Note on the Parallel Diffusion Coefficient}.
\newblock \emph{\apj}, 162:\penalty0 1049, December 1970.

\bibitem[{Hockney} and {Eastwood}(1989)]{1989ComSimParHockney}
R.~W. {Hockney} and J.~W. {Eastwood}.
\newblock {Computer simulation using particles}.
\newblock \emph{CRC Press}, 1989.

\bibitem[{Jokipii}(1966)]{1966ApJ...146..480J}
J.~R. {Jokipii}.
\newblock {Cosmic-Ray Propagation. I. Charged Particles in Random Magnetic
  Field}.
\newblock \emph{\apj}, 146:\penalty0 480, November 1966.

\bibitem[{Jokipii}(1987)]{1987ApJ...313..842J}
J.~R. {Jokipii}.
\newblock {Rate of energy gain and maximum energy in diffusive shock
  acceleration}.
\newblock \emph{\apj}, 313:\penalty0 842--846, February 1987.

\bibitem[{Jones} and {Ellison}(1991)]{1991SSRv...58..259J}
F.~C. {Jones} and D.~C. {Ellison}.
\newblock {The plasma physics of shock acceleration}.
\newblock \emph{\ssr}, 58:\penalty0 259--346, December 1991.

\bibitem[{Kloeden} et~al.(1997){Kloeden}, {Platen}, and
  {Schurz}]{1997NumSolSDE}
P.~E. {Kloeden}, E.~P. {Platen}, and H.~{Schurz}.
\newblock {Numerical solution of SDE through computer experiments}.
\newblock \emph{Springer-Verlag}, 1997.

\bibitem[{Krymskii}(1977)]{1977DoSSR.234.1306K}
G.~F. {Krymskii}.
\newblock {A regular mechanism for the acceleration of charged particles on the
  front of a shock wave}.
\newblock \emph{Dokl. Akad. Nauk SSSR}, 234:\penalty0 1306--1308, June 1977.

\bibitem[{Li} et~al.(2003){Li}, {Zank}, and {Rice}]{2003JGRA..108.1082L}
G.~{Li}, G.~P. {Zank}, and W.~K.~M. {Rice}.
\newblock {Energetic particle acceleration and transport at coronal mass
  ejection-driven shocks}.
\newblock \emph{\jgr}, 108:\penalty0 1082, February 2003.

\bibitem[{Li} et~al.(2005){Li}, {Zank}, and {Rice}]{2005JGRA..110.6104L}
G.~{Li}, G.~P. {Zank}, and W.~K.~M. {Rice}.
\newblock {Acceleration and transport of heavy ions at coronal mass
  ejection-driven shocks}.
\newblock \emph{\jgr}, 110:\penalty0 A06104, June 2005.

\bibitem[{Li} et~al.(2008){Li}, {Hu}, {Verkhoglyadova}, {Zank}, {Lin}, and
  {Luhmann}]{2008AIPC.1039.....L}
G.~{Li}, Q.~{Hu}, O.~{Verkhoglyadova}, G.~P. {Zank}, R.~P. {Lin}, and
  J.~{Luhmann}, editors.
\newblock \emph{{Particle acceleration and transport in the heliosphere and
  beyond:7th Annual International Astrophysics Conference}}, volume 1039 of
  \emph{American Institute of Physics Conference Series}, October 2008.

\bibitem[{Mann} et~al.(2003){Mann}, {Klassen}, {Aurass}, and
  {Classen}]{2003A&A...400..329M}
G.~{Mann}, A.~{Klassen}, H.~{Aurass}, and H.-T. {Classen}.
\newblock {Formation and development of shock waves in the solar corona and the
  near-Sun interplanetary space}.
\newblock \emph{\aap}, 400:\penalty0 329--336, March 2003.

\bibitem[{Mason} et~al.(2012){Mason}, {Li}, {Cohen}, {Desai}, {Haggerty},
  {Leske}, {Mewaldt}, and {Zank}]{2012ApJ...761..104M}
G.~M. {Mason}, G.~{Li}, C.~M.~S. {Cohen}, M.~I. {Desai}, D.~K. {Haggerty},
  R.~A. {Leske}, R.~A. {Mewaldt}, and G.~P. {Zank}.
\newblock {Interplanetary Propagation of Solar Energetic Particle Heavy Ions
  Observed at 1 AU and the Role of Energy Scaling}.
\newblock \emph{\apj}, 761:\penalty0 104, December 2012.

\bibitem[{Ng} and {Reames}(2008)]{2008ApJ...686L.123N}
C.~K. {Ng} and D.~V. {Reames}.
\newblock {Shock Acceleration of Solar Energetic Protons: The First 10
  Minutes}.
\newblock \emph{\apjl}, 686:\penalty0 L123--L126, October 2008.

\bibitem[{Ng} et~al.(1999){Ng}, {Reames}, and {Tylka}]{1999GeoRL..26.2145N}
C.~K. {Ng}, D.~V. {Reames}, and A.~J. {Tylka}.
\newblock {Effect of proton-amplified waves on the evolution of solar energetic
  particle composition in gradual events}.
\newblock \emph{\grl}, 26:\penalty0 2145--2148, 1999.

\bibitem[{Ng} et~al.(2003){Ng}, {Reames}, and {Tylka}]{2003ApJ...591..461N}
C.~K. {Ng}, D.~V. {Reames}, and A.~J. {Tylka}.
\newblock {Modeling Shock-accelerated Solar Energetic Particles Coupled to
  Interplanetary Alfv{\'e}n Waves}.
\newblock \emph{\apj}, 591:\penalty0 461--485, July 2003.

\bibitem[{Reames} and {Ng}(1998)]{1998ApJ...504.1002R}
D.~V. {Reames} and C.~K. {Ng}.
\newblock {Streaming-limited Intensities of Solar Energetic Particles}.
\newblock \emph{\apj}, 504:\penalty0 1002, September 1998.

\bibitem[{Sandroos} and {Vainio}(2006)]{2006A&A...455..685S}
A.~{Sandroos} and R.~{Vainio}.
\newblock {Particle acceleration at shocks propagating in inhomogeneous
  magnetic fields}.
\newblock \emph{\aap}, 455:\penalty0 685--695, August 2006.

\bibitem[{Sandroos} and {Vainio}(2007)]{2007ApJ...662L.127S}
A.~{Sandroos} and R.~{Vainio}.
\newblock {Simulation Results for Heavy Ion Spectral Variability in Large
  Gradual Solar Energetic Particle Events}.
\newblock \emph{\apjl}, 662:\penalty0 L127--L130, June 2007.

\bibitem[{Schlickeiser} et~al.(2002){Schlickeiser}, {Vainio}, {B{\"o}ttcher},
  {Lerche}, {Pohl}, and {Schuster}]{2002A&A...393...69S}
R.~{Schlickeiser}, R.~{Vainio}, M.~{B{\"o}ttcher}, I.~{Lerche}, M.~{Pohl}, and
  C.~{Schuster}.
\newblock {Conversion of relativistic pair energy into radiation in the jets of
  active galactic nuclei}.
\newblock \emph{\aap}, 393:\penalty0 69--87, October 2002.

\bibitem[{Torsti} et~al.(1996){Torsti}, {Kocharov}, {Vainio}, {Anttila}, and
  {Kovaltsov}]{1996SoPh..166..135T}
J.~{Torsti}, L.~G. {Kocharov}, R.~{Vainio}, A.~{Anttila}, and G.~A.
  {Kovaltsov}.
\newblock {The 1990 May 24 Solar Cosmic-Ray Event}.
\newblock \emph{\solphys}, 166:\penalty0 135--158, June 1996.

\bibitem[{Tylka} and {Lee}(2006)]{2006ApJ...646.1319T}
A.~J. {Tylka} and M.~A. {Lee}.
\newblock {A Model for Spectral and Compositional Variability at High Energies
  in Large, Gradual Solar Particle Events}.
\newblock \emph{\apj}, 646:\penalty0 1319--1334, August 2006.

\bibitem[{Vainio} and {Laitinen}(2007)]{2007ApJ...658..622V}
R.~{Vainio} and T.~{Laitinen}.
\newblock {Monte Carlo Simulations of Coronal Diffusive Shock Acceleration in
  Self-generated Turbulence}.
\newblock \emph{\apj}, 658:\penalty0 622--630, March 2007.

\bibitem[{Vainio} and {Schlickeiser}(1998)]{1998A&A...331..793V}
R.~{Vainio} and R.~{Schlickeiser}.
\newblock {Alfven wave transmission and particle acceleration in parallel shock
  waves}.
\newblock \emph{\aap}, 331:\penalty0 793--799, March 1998.

\bibitem[{Vainio} et~al.(2000){Vainio}, {Kocharov}, and
  {Laitinen}]{2000ApJ...528.1015V}
R.~{Vainio}, L.~{Kocharov}, and T.~{Laitinen}.
\newblock {Interplanetary and Interacting Protons Accelerated in a Parallel
  Shock Wave}.
\newblock \emph{\apj}, 528:\penalty0 1015--1025, January 2000.

\bibitem[{Webb}(1983)]{1983ApJ...270..319W}
G.~M. {Webb}.
\newblock {First order and second order Fermi acceleration of energetic charged
  particles by shock waves}.
\newblock \emph{\apj}, 270:\penalty0 319--338, July 1983.

\bibitem[{Webb} et~al.(1983){Webb}, {Axford}, and
  {Terasawa}]{1983ApJ...270..537W}
G.~M. {Webb}, W.~I. {Axford}, and T.~{Terasawa}.
\newblock {On the drift mechanism for energetic charged particles at shocks}.
\newblock \emph{\apj}, 270:\penalty0 537--553, July 1983.

\bibitem[{Whitham}(1965)]{1965JFM....22..273W}
G.~B. {Whitham}.
\newblock {A general approach to linear and non-linear dispersive waves using a
  Lagrangian}.
\newblock \emph{\jfm}, 22:\penalty0 273--283, 1965.

\bibitem[{Zank} et~al.(2000){Zank}, {Rice}, and {Wu}]{2000JGR...10525079Z}
G.~P. {Zank}, W.~K.~M. {Rice}, and C.~C. {Wu}.
\newblock {Particle acceleration and coronal mass ejection driven shocks: A
  theoretical model}.
\newblock \emph{\jgr}, 105:\penalty0 25079--25096, November 2000.

\end{thebibliography}

\end{document}